\documentclass[fleqn,usenatbib]{mnras}

\usepackage{newtxtext,newtxmath}

\usepackage[T1]{fontenc}
\usepackage{ae,aecompl}

\usepackage{graphicx}	
\usepackage{amsmath}	
\usepackage{amssymb}	

\title[The nuclear region of NGC 613]{The nuclear region of NGC 613. I - Multiwavelength analysis}

\author[Patr\'icia da Silva et al.]{
Patr\'icia da Silva$^1$ \thanks{\href{mailto:p.silva2201@gmail.com}{p.silva2201@gmail.com}},
R. B. Menezes$^{2}$ \thanks{\href{mailto:roberto.menezes@maua.br}{roberto.menezes@maua.br}},
J. E. Steiner$^1$ \thanks{\href{mailto:joao.steiner@iag.usp.br}{joao.steiner@iag.usp.br}}
\\
$^1$Instituto de Astronomia, Geof\'isica e Ci\^encias Atmosf\'ericas, Departamento de Astronomia, Universidade de S\~ao Paulo, 05508-090, S\~ao Paulo, SP, Brazil\\
$^2$Instituto Mau\'a de Tecnologia, Pra\c{c}a Mau\'a 1, 09580-900, S\~ao Caetano do Sul, SP, Brazil  
}
\date{Accepted 2019 December 28. Received 2019 November 22; in original form 2019 August 16.}

\pubyear{2020}

\begin{document}
\label{firstpage}
\pagerange{\pageref{firstpage}--\pageref{lastpage}}
\maketitle

\begin{abstract}

 In this paper, we report a detailed study with a variety of data from optical, near-infrared, X-ray, and radio telescopes of the nuclear region of the galaxy NGC 613 with the aim of understanding its complexity. We detected an extended stellar emission in the nucleus that, at first, appears to be, in the optical band, two stellar nuclei separated by a stream of dust. The active galactic nucleus (AGN) is identified as a variable point-like source between these two stellar components. There is a central hard X-ray emission and an extended soft X-ray emission that closely coincides with the ionization cone, as seen in the [O \textsc{iii}]$\lambda$5007 emission. The centroid of the [O \textsc{i}]$\lambda$6300 emission does not coincide with the AGN, being shifted by 0.24 arcsec towards the ionization cone; this shift is probably caused by a combination of differential dust extinction together with emission and reflection in the ionization cone. The optical spectra extracted from the central region are typical of low-ionization nuclear emission-line regions. We also identify 10 H \textsc{ii} regions, eight of them in a star forming ring that is visible in Br$\gamma$, [Fe \textsc{ii}]$\lambda$16436 and molecular CO(3-2) images observed in previous studies. Such a ring also presents weak hard X-ray emission, probably associated with supernova remnants, not detected in other studies. The position of the AGN coincides with the centre of a nuclear spiral (detected in previous works) that brings gas and dust from the bar to the nucleus, causing the high extinction in this area.

\end{abstract}




\section{Introduction}\label{introducao}

Active galactic nuclei (AGNs) are phenomena in which a significant amount of energy is produced by a non-stellar source. Such energy is believed to be produced by gas accretion onto a supermassive black hole  (\citealt{lyndenbell}; see \citealt{netzer} for more details). The study of the environment around AGNs is relevant for the evaluation of their interaction with the rest of the host galaxy through the processes of feeding and feedback (\citealt{ngc4151,ngc4151_2}). The feeding process can be studied by observations of molecular emission in the near-infrared (NIR) or millimetre bands. The feedback can be studied, for instance, by observing optical emission lines associated with outflows \citep{outflows} and ionization cones or by observing radio jets \citep{Beall_2003}. The emission of the AGN itself can be observed from radio to $\gamma$-rays. By analysing multiwavelength data it is possible to disentangle the complexity of the AGN, its environment, and their interaction.

Statistical properties of AGNs are well derived from studies of large samples (e.g. \citealt{ho2008}; \citealt{gultekin}). However, the interaction of an AGN with its environment can be quite complex and individual objects must be studied in detail if we want to understand this complexity. 

In this work, we present a multiwavelength study of the nuclear region of NGC 613. This is an SB(rs)bc galaxy, located at 26 $\pm$ 5 Mpc \citep{distancia}, and is known to have a composite nucleus \citep{veron} with an AGN that interacts with the galaxy through outflows, an ionization cone, and a radio jet, as well as a circumnuclear ring of H \textsc{ii} regions that has a radius of about 300 pc (\citealt{hummel}, \citealt{falcon613} and \citealt{audibert}).

\begin{figure}
\begin{center}
 \includegraphics[scale=0.5]{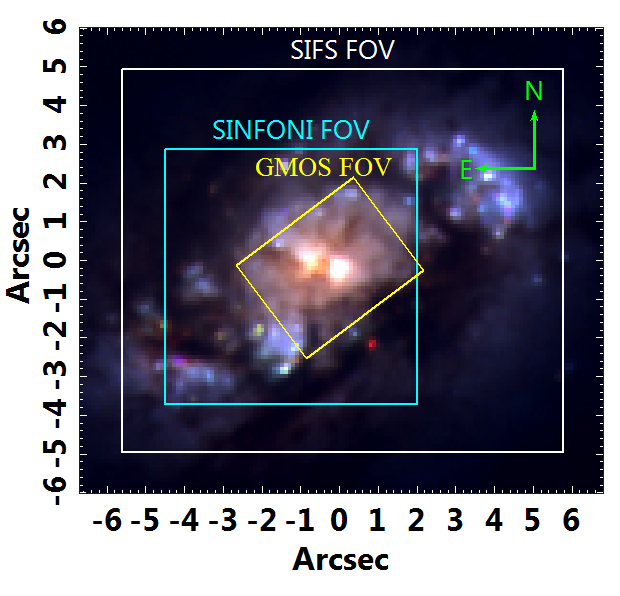}
  \caption{RGB composition of the images in the filters \textit{F 814 W} (red), \textit{F 606 W} (green), and \textit{F 450 W} (blue) of \textit{HST}/WFPC2. The positions and sizes of each field of view (FOV) of GMOS, SINFONI, and SIFS data cubes are represented by the squares, with the orientation N-E. The PA of the GMOS observation is 127$\degr$. The total size of the FOV of this image is 13.55 arcsec $\times$ 12.05 arcsec. \label{hst_tot}}
\end{center}
\end{figure}

That the galaxy contains an AGN is confirmed, for example, by the detection of high-ionization emission lines, such as [Ne \textsc{v}] and [O \textsc{iv}], observed with the \textit{Spitzer} space telescope \citep{goulding}. Besides that, a radio jet, showing a structure that extends along 5 arcsec with position angle (PA) of 6$^{\circ}$ (almost perpendicular to the bar) was observed with the Very Large Array (VLA) at 6 and 20 cm \citep{hummel2}. At 4.86 and 14.94 GHz, three blobs were detected in the central kpc, forming a linear structure whose PA is 12$^{\circ}$ \citep{hummel}. \citet{miyamoto1} also detected  an elongated structure in 95 GHz with PA= 20$^{\circ}$ $\pm$ 8$^{\circ}$. All these data show the presence of a nuclear radio jet. Associated with this jet, there is an anisotropic maser with luminosity of $\sim$ 15.9 $L_{\bigodot}$ \citep{kondratko}. There is also a megamaser (in the same position as the previous one) with isotropic emission and luminosity of $\sim$ 35 $L_{\bigodot}$, which can also be associated with the nuclear jet \citep{castangia1}. 

The AGN in NGC 613 is of low luminosity. Using an X-ray luminosity of L$_{2-10 kev} =8.92 \times 10^{40}$ erg s$^{-1}$ \citep{castangia2013} and a bolometric correction of 20 \citep{vasudevan}, \citet{daviesbecca} determined that the AGN bolometric luminosity is 1.6 $\times 10^{42}$ erg s$^{-1}$. From the analysis of the X-ray Multi-Mirror Mission (\textit{XMM-Newton}) data, \citet{castangia2013} obtained an intrinsic column density of 36$^{+5}_{-4} \times 10^{22} cm^{-2}$ and an emission ratio of soft X-rays (2-10 keV) to [O \textsc{iii}]$\lambda$5007 of 5.6 for NGC 613 centre. \citet{ASMUS2015} determined a column density  of 23.5$\pm$ 0.5 $\times 10^{22} cm^{-2}$, confirming the highly obscured AGN.

The Atacama Large Millimeter/Submillimeter Array (ALMA) observation of molecular lines revealed that the nucleus has emission compatible with the ones of Seyfert/LINER composite and is also being ionized by shock heating. There are very energetic molecular outflows that are possibly fossils from a phase when this AGN was in a higher activity and they are also associated with the radio jet, possibly being driven by it \citep{audibert}.

Through the decomposition of gas ionization mechanisms of the central $\sim$ 35 arcsec $\times$ 25 arcsec, by using a spectral basis that consider ionization by shock, AGN, and stars, an ionization cone aligned with the radio jet was detected \citep{daviesbecca}. There are also shock waves in the ionization cone edges (inside the central 1 kpc$^2$), probably formed by outflows of gas coming from the AGN that are shocking gas in the interstellar medium.

It is believed that circumnuclear star-forming rings are the result of the strong interaction between the bar (or strong spiral arms) and the inner gas located in the Lindblad resonance region \citep{elmegreen}.  According to \citet{BOKERIAU,boker} and \citet{falconIAU,falcon613}, the circumnuclear ring of NGC 613 is formed by seven H \textsc{ii} regions. This ring can also be observed in radio wavelengths (\citealt{hummel}; \citealt{miyamoto1, miyamoto2}; \citealt{audibert}) and, if its structure is circular, the inclination should be 55$^{\circ}$ $\pm$ 5$^{\circ}$ \citep{hummel}. \citet{Combess} and \citet{audibert} found that this ring is connected to the bar in two points, at NW and SE, as \citet{boker} previously estimated by using the \textit{Hubble Space Telescope (HST)} images, and that the ring is indeed clumpy as it is shown in the image of Br$\gamma$ \citep{falcon613}. There is also a nuclear spiral of molecular gas connected with the ring in two different spots \citep{audibert}.

In this article, we analyse the central region of NGC 613 using data cubes obtained with the Gemini Multi-Object Spectrograph (GMOS) from the Gemini-South telescope and  Soar Integral Field Spectrograph (SIFS) from the SOAR telescope and the archived data from the \textit{HST}, from ALMA, from the Spectrograph for Integral Field Observations in the Near Infrared (SINFONI) of the Very Large Telescope (VLT), and from the \textit{Chandra} Space Telescope. The purpose of our analysis is to correlate our results with the information and data from the literature in order to explain the nature of the observed emission in the centre of this galaxy. This paper (Paper I) is part of a comprehensive analysis of NGC 613 nucleus and presents the study of the emission of this galaxy centre. The study of the stellar and gas kinematics and stellar archaeology will be presented in Paper II. 

Section \ref{secobs} describes the observations and treatment methods of the data obtained with the many instruments used in this work. Section \ref{emissao} shows the optical emission-line analysis of the observed regions in the centre of NGC 613. The study of X-ray emission is presented in section \ref{secraiosx}. We discuss the data in section \ref{secdiscussion}, presenting the possible scenarios, and section \ref{secconclusion} summarizes the conclusions of this work. In Appendix \ref{espectros_analise}, we present the optical extracted spectra of the observed regions and the method that we used to calculate  the integrated flux of the emission lines, in order to calculate their ratios. 

\section{Observations and data reduction}\label{secobs}

This work presents data obtained with different instruments: Wide-Field Planetary Camera 2 (WFPC2) from \textit{HST}, GMOS from Gemini-South telescope, SINFONI from VLT, SIFS from SOAR telescope, Advanced CCD Imaging Spectrometer (ACIS) from \textit{Chandra} space telescope and ALMA. The following subsections describe the treatment and conditions of each observation.

Fig. \ref{hst_tot} shows the image of NGC 613 nucleus, obtained with \textit{HST}, and the fields of view (FOVs) of the GMOS, SINFONI, and SIFS data cubes. The GMOS data cube has PA = 127$^{\circ}$, so its orientation is different from the other FOVs, whose PAs = 0$^{\circ}$. By comparing the three data cubes (optical and NIR), the SIFS FOV is the largest, followed by SINFONI (which has the highest spatial resolution) and by GMOS.

\begin{figure*}
\begin{center}
 \includegraphics[scale=0.368]{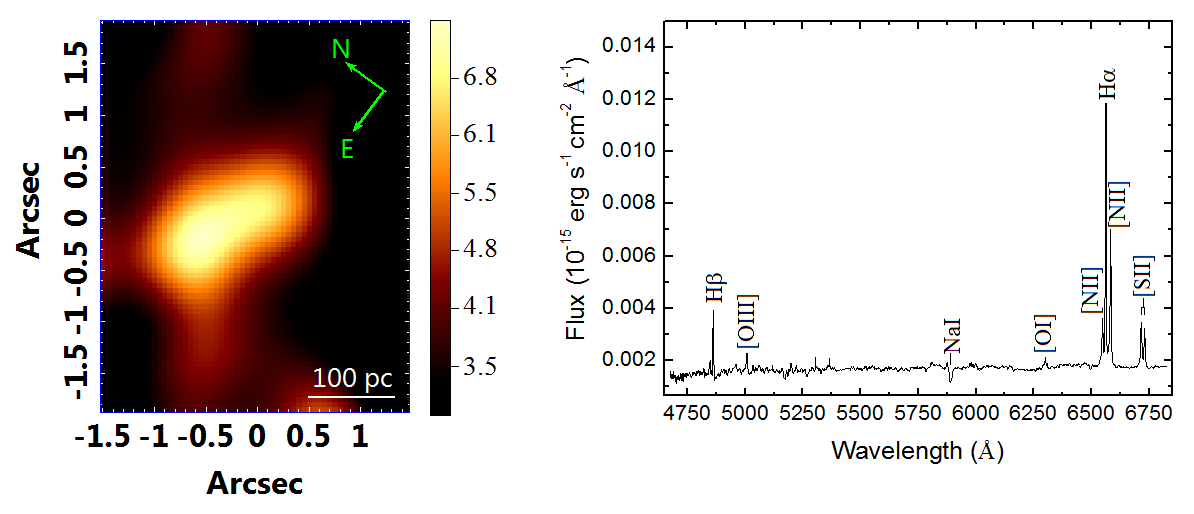}
  \caption{Image of the GMOS data cube, obtained after the treatment, of the central region of NGC 613 collapsed along the spectral axis with its flux scale and with its average spectrum corrected for redshift. Note the indication of N-E and the scale of 100 pc. \label{gmoscolapsed}}
\end{center}
\end{figure*}

\begin{figure*}
\begin{center}
 \includegraphics[scale=0.38]{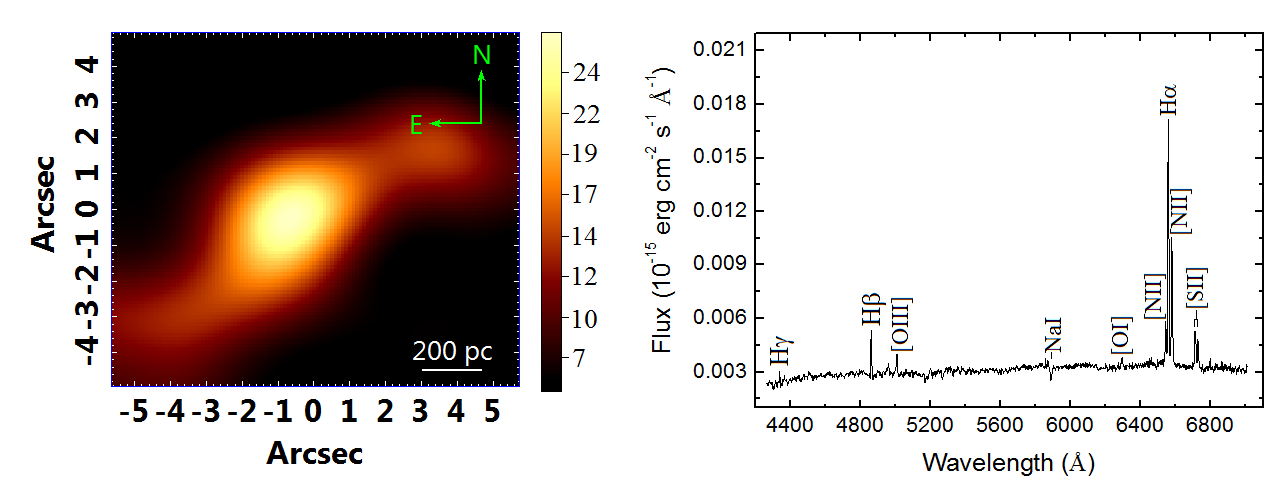}
  \caption{Image of the SIFS data cube, obtained after the treatment, of the central region of NGC 613 collapsed along the spectral axis with its flux scale and with its average spectrum corrected for redshift. Note the indication of N-E and the scale of 200 pc. \label{sifscolapsed}}
\end{center}
\end{figure*}

\begin{figure*}
\begin{center}
 \includegraphics[scale=0.38]{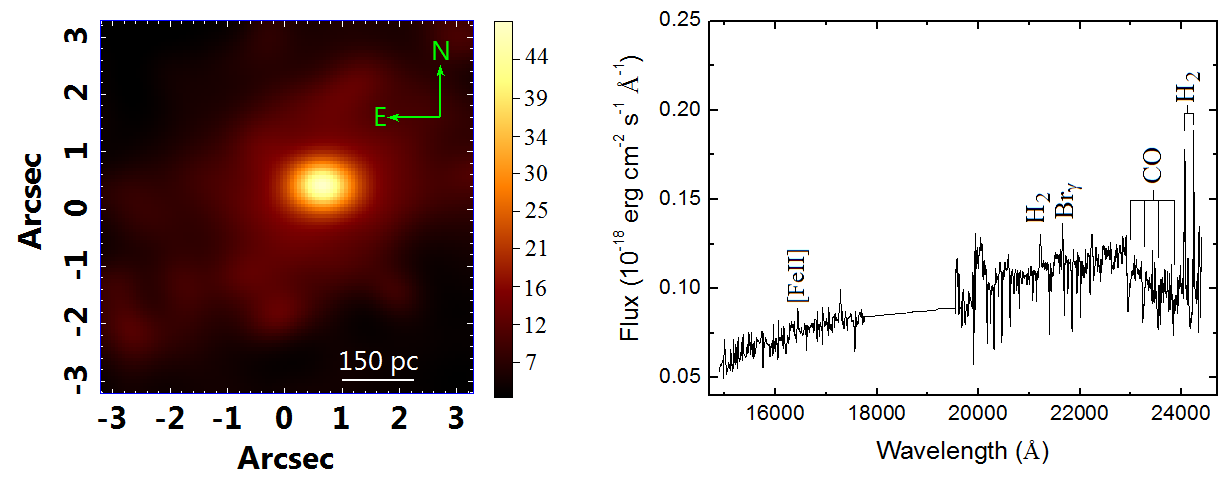}
  \caption{Image of the SINFONI data cube, obtained after the treatment, of the central region of NGC 613 collapsed along the spectral axis with its flux scale and with its average spectrum corrected for redshift. Note the indication of N-E and the scale of 150 pc. \label{sinfonicolapsed}}
\end{center}
\end{figure*}

\subsection{\textit{Hubble Space Telescope} images}\label{hstdata}

 In this study, we used images obtained from the \textit{HST} public archive. Those images were part of the observation programs 9042 (PI: Smartt, S. J.) and 15133 (PI: Erwin, P.) , obtained with the WFPC2 instrument on 2001 August 21 and WFPC3 on 2018 August 22. For the program 9042, the images in the filters \textit{F 450 W}, \textit{F 606 W} and \textit{F 814 W} were taken with an integration time of 160 s. Fig. \ref{hst_tot} shows the FOV in the central area of the galaxy, whose size is 13.6 arcsec $\times$ 12.1 arcsec, in those observed filters. On the other hand, for the program 15133, the integration time for the image of \textit{F 814 W} filter was 500 s and for the image of \textit{F 475 W} filter was 700 s. 

\subsection{Optical data cubes}

Although having similar spectral range, the data cubes obtained with the integral field units (IFUs) of GMOS and SIFS present some significant differences. The SIFS FOV is almost 6.7 times larger than the GMOS FOV, in area, which allows us to analyse the circumnuclear region of this galaxy on a larger scale. In order to evaluate the signal-to-noise ratio (S/N) of these data cubes, we extracted a spectrum of a rectangular area with 0.3 arcsec $\times$ 0.3 arcsec (equivalent to 1 spaxel of a raw data cube obtained with SIFS and a square of 6 spaxels (spatial pixels) $\times$ 6 spaxels of one of the raw GMOS data cubes), centred in the brightest point visible in the data cubes. Then, we calculated the S/N ratio in each spectrum, in the spectral range from 5610 to 5700 \AA. The results were  $S/N_{GMOS}$ = 29 and $S/N_{SIFS}$ = 27. Considering that and also taking into account the exposure times for the GMOS data (930 s) and for the SIFS data (900 s), we conclude that these two instruments are very similar, regarding the S/N ratio. As discussed in the following sections, the difference between the spatial resolution of those data cubes is highly significant, since the seeing of the observation taken with SIFS was considerably larger than the seeing of the GMOS observation, resulting in the full width at half-maximum (FWHM) of the point spread function (PSF) in the SIFS data cube being three times larger than the FWHM of the PSF obtained with the GMOS data cube. 

\subsubsection{GMOS data}

These data were taken using the IFU of the GMOS, in one-slit mode. The data are part of the \textit{Deep IFS View of Nuclei of Galaxies} (DIVING3D) survey, which aims to analyse the nuclear regions of all Southern-hemisphere galaxies brighter than B=12 (Steiner, J. E. et al., in preparation). 

The observations of NGC 613 were taken on 2015 January 25 as part of the observation program GS-2014B-Q-30 in the Gemini-South telescope. Three 930 s exposures were taken with spatial dithering and PA = 127$^{\circ}$. Using the grating R831+G5322, centred in 5850 \AA, we obtained a spectral resolution of R=4340 and a spectral coverage from 4675 to 6828 \AA.

We used data reduction packages, developed by the Gemini observatory, in \textsc{iraf} environment. This reduction was conducted using the following processes: trim determination, bias subtraction, cosmic ray removal (\textsc{lacos} -- \citealt{van01}), spectra extraction, corrections of gain variations between the spectral pixels (using GCAL-flat images), corrections of gain variations between the spaxels (using twilight-flat images), wavelength calibration (using images of the CuAr lamp),  subtraction of the average spectrum of the FOV corresponding to the sky observation (located at 1 arcmin from the object), flux calibration, atmospheric extinction correction, telluric absorption removal, and data cube construction. The resulting three data cubes had spaxels of 0.05 arcsec $\times$ 0.05 arcsec and FOV of 3.5 arcsec $\times$ 5 arcsec. The FWHM of the PSF calculated from the acquisition image (0.72 arcsec), in this case, is compatible with the FWHM calculated from the image of [O \textsc{i}]$\lambda$6300 from the data cube after the reduction (0.71 arcsec). So we used the last one (0.71 arcsec) in the process of deconvolution described in section \ref{datacubetreatment}.

Fig. \ref{gmoscolapsed} shows the GMOS data cube after the treatment (see section \ref{datacubetreatment}) collapsed along the spectral axis, with its average spectrum. We can see an extended source in the central region and the main optical emission lines, quite broad (see Figs. \ref{espectrosN1N2} and \ref{decomposicaoN1N2} of Appendix \ref{espectros_analise}), indicating high-velocity gas emission.

\subsubsection{SIFS data}

The data were taken on 2017 November 23 with SIFS on the SOAR telescope, during the Science Verification program. Six 900 s exposures were taken with spatial dithering, in order to obtain a mosaic of the central region of NGC 613. Three of these exposures were centred 3 arcsec north of the nucleus of the galaxy and the other three were centred 3 arcsec south of the nucleus. The dither step for each of these groups of three exposures was 0.3 arcsec. The PA of all the observations was 0\degr. The plate scale was 0.3 arcsec/fibre and the grating used was 700 l/mm, centred in 5650 \AA, which resulted in a spectral coverage from 4250 to 7050 \AA\ and spectral resolution of R = 4200. During the same night, data from the standard star LTT 2415 were taken with exposure time of 300 s. Since SIFS has an Atmospheric Dispersion Corrector (ADC), the differential atmospheric refraction effect was removed from the observed data.   

The reduction was  performed using scripts developed in Iteractive Data Language (\textsc{idl}) and included the following steps: correction of dead fibres, spectra extraction, flat-field corrections (in order to remove gain variations between the spectral pixels and between fibres), wavelength calibration (using images of an HgAr lamp), sky subtraction, and creation of data cubes. After this, processes of flux calibration, atmospheric extinction correction, and telluric absorption removal were applied using scripts developed in \textsc{iraf} environment. At the end, we constructed a mosaic with the six reduced data cubes and obtained a final data cube with spaxels of 0.3 arcsec $\times$ 0.3 arcsec and FOV with 15 arcsec $\times$ 7.8 arcsec.

Since the standard star was observed in the same night, we estimated the FWHM of the PSF of this observation from its data cube, and the result was 2.4 arcsec. 

After that, the data cube treatment was applied as described in section \ref{datacubetreatment}. Fig. \ref{sifscolapsed} shows the image of the treated data cube, collapsed along the spectral axis, and its average spectrum, with the optical main emission lines indicated. Note a central source with extended emission in the SE and NW directions.

\subsection{SINFONI data}

The data were obtained from the public archive and were taken on 2005 October 23 and on 2005 November 13  as part of the observation program 076.B-0646(A), PI: B\"{o}ker,T., with four and five exposures of 300 s, respectively, in the H+K band, which resulted in a spectral resolution of R = 1500 and a spectral range from 15000 to 24500 \AA. The PA of the observation was 0$^{\circ}$ with fore-optics of  0.25 arcsec. This data set was already published by \citet{boker} and \citet{falcon613}. Here, we use these data superposed to SIFS data to locate the H \textsc{ii} regions, in order to calculate the emission-line ratios in the SIFS data cube and to make some comparisons with the other data. In paper II, we use these data to study the stellar and gas kinematics.

The reduction of the data was performed using the \textsc{gasgano} software in a process involving the following steps: bad pixel correction, flat-field correction, spatial rectification in order to remove potential distortions along the FOV, wavelength calibration (using Ne lamp images), sky subtraction, and data cube construction. The flux calibration and telluric absorption removal were applied later, using scripts developed in \textsc{iraf} environment. We obtained nine data cubes with spaxels of 0.125 arcsec $\times$ 0.125 arcsec and FOV of 8 arcsec $\times$ 8 arcsec.  

The FWHM of the PSF of the data cube before treatment, estimated from the image of H$_2\lambda21218$, was 0.69 arcsec.

Although the data reduction procedure we used is analogous to the one adopted by \citet{boker} and \citet{falcon613}, we also applied a treatment procedure to the reduced data cubes (not used by the aforementioned authors), in order to obtain certain improvements, as described in section \ref{datacubetreatment}. Fig. \ref{sinfonicolapsed} shows the image of the data cube collapsed along the spectral axis and the data cube average spectrum, showing the main emission lines in this spectral range. 
 
\subsection{Optical and NIR data cubes treatment}\label{datacubetreatment}

The data cube treatment was applied using scripts written in \textit{idl} developed by our group (see \citealt{rob1,rob2,gmostreat}). We started by  performing the differential atmospheric refraction correction in the data cubes (except in the SIFS data cube that has ADC). After that, the data cubes were combined in only one: we combined the three GMOS data cubes in form of a median. As said previously, we had, after the reduction, nine SINFONI data cubes; they were combined in order to minimise sky subtraction problems, as some of them had sky emission in excess and some others exaggerated sky subtraction. The combinations were done from 3 to 3 data cubes using medians (alternating between the dates of the observations) until we had only one final data cube.  

\begin{figure*}
\begin{center}
 \includegraphics[scale=0.31]{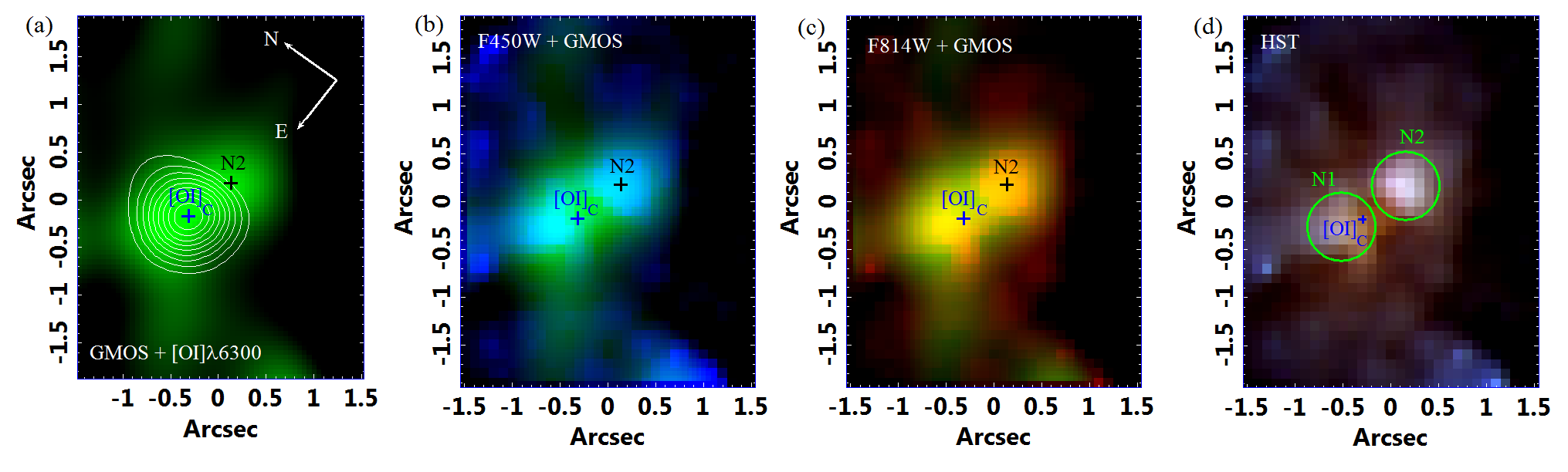}
  \caption{(a) Image of the GMOS data cube integrated along the spectral axis. The white contours represent the emission of [O \textsc{i}]$\lambda$6300 and its centre is indicated as [O \textsc{i}]$_{C}$ ($X_C= -0.35$ arcsec and $Y_C= -0.2$ arcsec). The position of N2 was defined as the centre of the emission northwest of N1 ($X_C= 0.1$ arcsec and $Y_C= 0.15$ arcsec). (b) GB composition of the image in the \textit{HST} filter \textit{F 450 W} (blue) and the image of the integrated GMOS data cube (green). (c) RG composition of the same GMOS data cube image (green) and the image in the \textit{HST} filter \textit{F 814 W} (red). (d) RGB composition in the \textit{HST} filters: \textit{F 814 W} (red), \textit{F 606 W} (green) and \textit{F 450 W} (blue). The circles indicate the regions N1 and N2. All the images have crosses that indicate the positions [O \textsc{i}]$_{C}$ and N2, whose sizes represent the uncertainty of 3$\sigma$, taking into account the size of the spaxels of the GMOS data cube. Since the last panel contains only the images from \textit{HST}, whose spaxels are larger than the GMOS spaxels, the size of the cross differs. \label{GMOS_HST}}
\end{center}
\end{figure*}

SINFONI and SIFS data cubes were spatially re-sampled to have spaxels of 0.0625 arcsec and 0.1 arcsec, respectively. All the data cubes were spatially filtered using the Butterworth method \citep{gwoods}, in order to remove the high spatial frequency noise. Then, the instrumental fingerprint removal was applied. Lastly, we applied the Richardson--Lucy deconvolution (\citealt{rich} and \citealt{lucy}). For more details about the Richardson--Lucy deconvolution and the other treatment techniques, see \citet{rob1,rob2,gmostreat}. In GMOS and SIFS data cubes, the PSF variation law was estimated using the standard stars data cubes that were used in the data cubes reduction processes. For the SINFONI data cube, we used a constant PSF. 

As said previously, the SIFS standard star was observed in the same night, so we used these data to estimate the FWHM of the PSF, at 5647 \AA, that was equal to 2.4 arcsec. The process of deconvolution was performed with six iterations. All the treatment was also performed in the standard star data cube, in order to obtain the FWHM of the PSF after the deconvolution process (1.6 arcsec).

Regarding the GMOS data cube, the best estimate of the FWHM of PSF was obtained from the [O \textsc{i}]$\lambda$6300 image. At first, we assumed that this emission is point-like. After applying the deconvolution, however, we noticed that the difference of the values of the FWHM of the PSF before and after of the deconvolution (0.71 arcsec and 0.68 arcsec) was not so significant as we usually get for other objects (we often get differences between 0.10 and 0.15 arcsec). This probably means that the emission of [O \textsc{i}]$\lambda$6300 is not point-like. Nevertheless, this is the most point-like emission in GMOS data cube. The number of iterations of the deconvolution process was 10. 

For the SINFONI data cubes, we had no other way to determine the PSF, only using an image of some emission line. In this case, we used the image of H$_2\lambda21218$, which presents a point-like emission in the inner centre of the FOV. The difference of the FWHM of the PSF before and after the deconvolution process was also lower than we expected. The deconvolution was applied with 10 iterations and the resulting FWHM of the PSF was  0.60 arcsec. It is important to mention that, as discussed in \citet{rob1, rob2,gmostreat}, the Richardson--Lucy deconvolution does not compromise the data in anyway (keeping the flux values and also the spatial morphology of the structures unchanged). Therefore, although the improvement in the spatial resolution, in this case, is not dramatic, such a technique is worth to be applied.

\begin{figure*}
\begin{center}
 \includegraphics[scale=0.325]{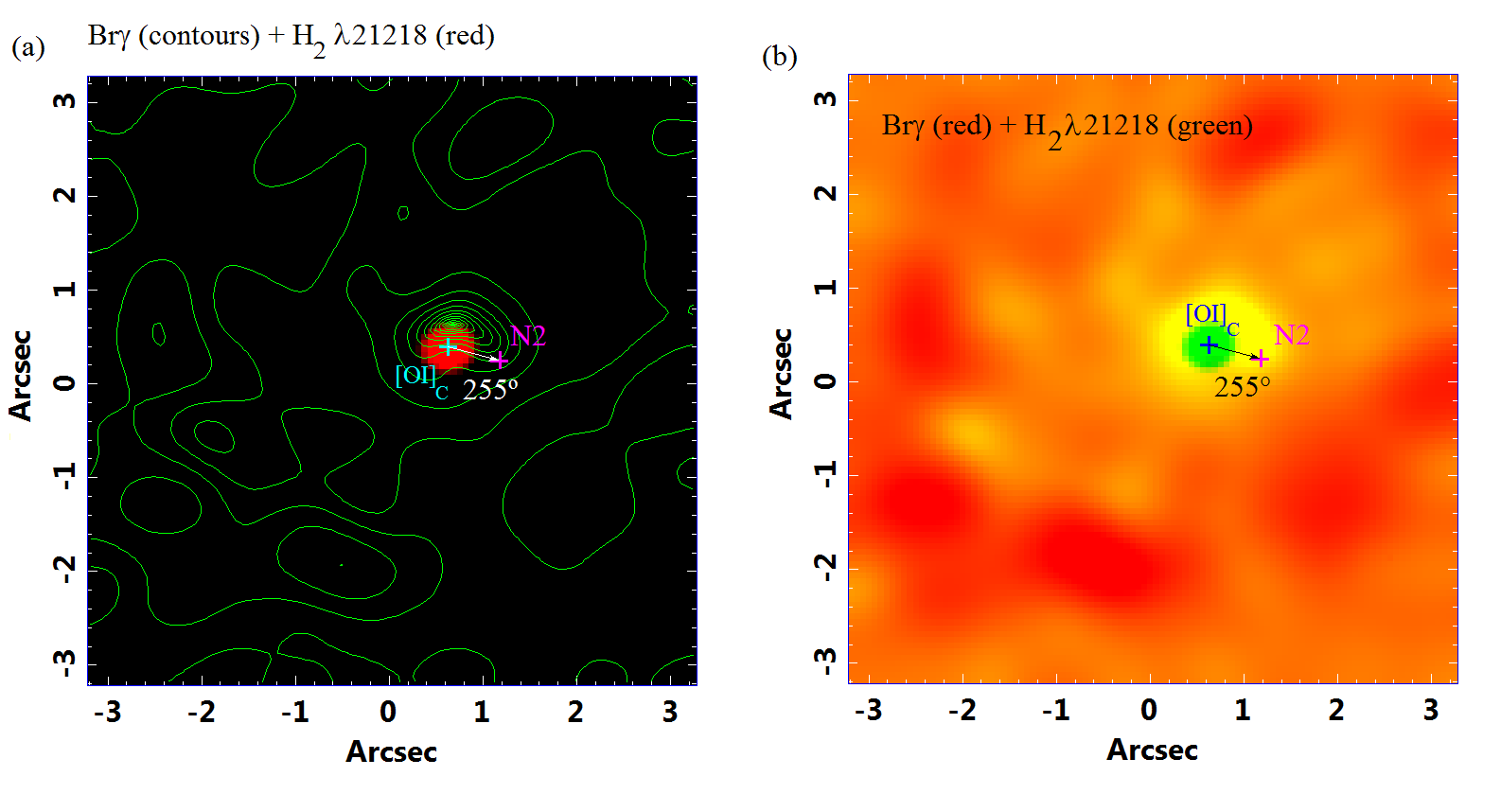}
  \caption{(a) Image of the centre emission of H$_2\lambda$21218 of the SINFONI gas data cube in red, with green contours of the Br$\gamma$ inverted image in flux units (or Br$\gamma$ in absorption). (b) RG composition of H$_2\lambda$21218 (green) and Br$\gamma$ (red). The yellow region is where Br$\gamma$ is in absorption or not being emitted. Both images have indications of the positions of [O \textsc{i}]$_{C}$ and of the centre of N2, assuming that the position of [O \textsc{i}]$_{C}$ is coincident with the peak of the H$_2\lambda$21218 emission and by propagating the position of N2, knowing the PA (255$^{\circ}$, indicated in the image) and distance between the [O \textsc{i}]$_{C}$ and N2 (centre of the brightest stellar nucleus from the \textit{HST} images).  \label{SINFONIN1N2}}
\end{center}
\end{figure*}

\begin{figure}
\begin{center}
   \includegraphics[scale=0.35]{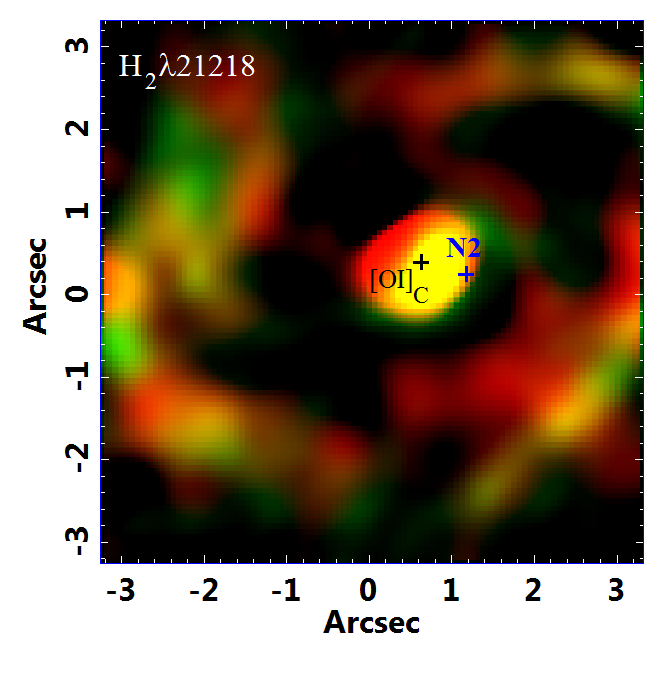}
  \caption{RG compositions with the CO(3-2) image obtained from ALMA data cube (in green) and the H$_2\lambda$21218 image obtained from SINFONI data cube (in red). The image shows the positions of [O \textsc{i}]$_{C}$ and of the centre of N2. The size of the crosses represents the 3$\sigma$ uncertainty taking into account the size of SINFONI spaxels. This image was made to match the SINFONI and ALMA data using the circumnuclear ring position and also to locate [O \textsc{i}]$_{C}$ and the centre of N2 in ALMA data. } \label{ANEL_NIR_ALMA}
\end{center}
\end{figure}

\begin{figure*}
\begin{center}
 \includegraphics[scale=0.325]{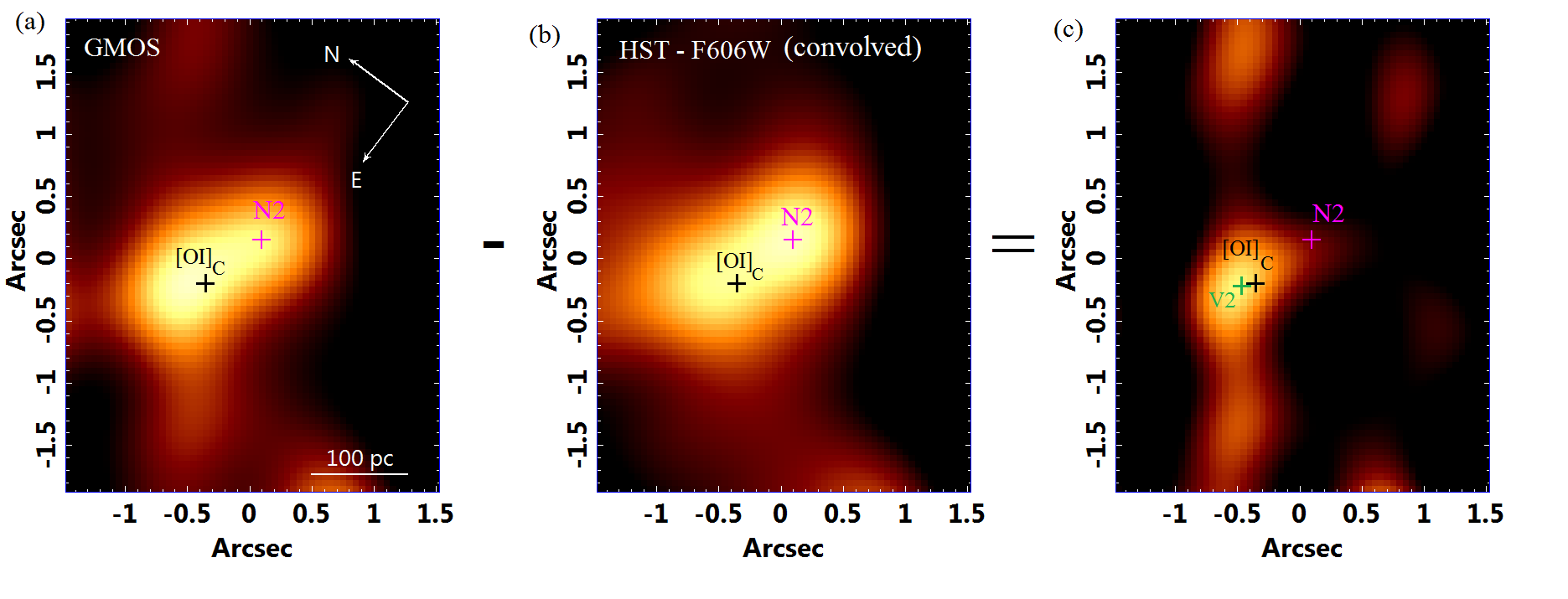}
  \caption{(a) GMOS data cube integrated image considering the transmission curve of the \textit{HST} filter \textit{F 606 W}. (b) Image in the \textit{HST} filter \textit{F 606 W} convolved with the PSF of the GMOS data cube (0.71 arcsec). Both (a) and (b) had their fluxes normalized at N2. One can note that, in the GMOS image, there is an emission excess near [O \textsc{i}]$_{C}$ and, in the \textit{HST} image, this excess is located at N2. The subtraction of these two images (c) shows a source near [O \textsc{i}]$_{C}$ (the green cross represents its centre), whose projected distance to [O \textsc{i}]$_{C}$ is  0.11 arcsec. We named this source v2. The sizes of the crosses represent the uncertainty of 3$\sigma$ for the positions of the sources. \label{GMOSMENOSHST}}
\end{center}
\end{figure*}

\begin{figure}
\begin{center}
 \includegraphics[scale=0.37]{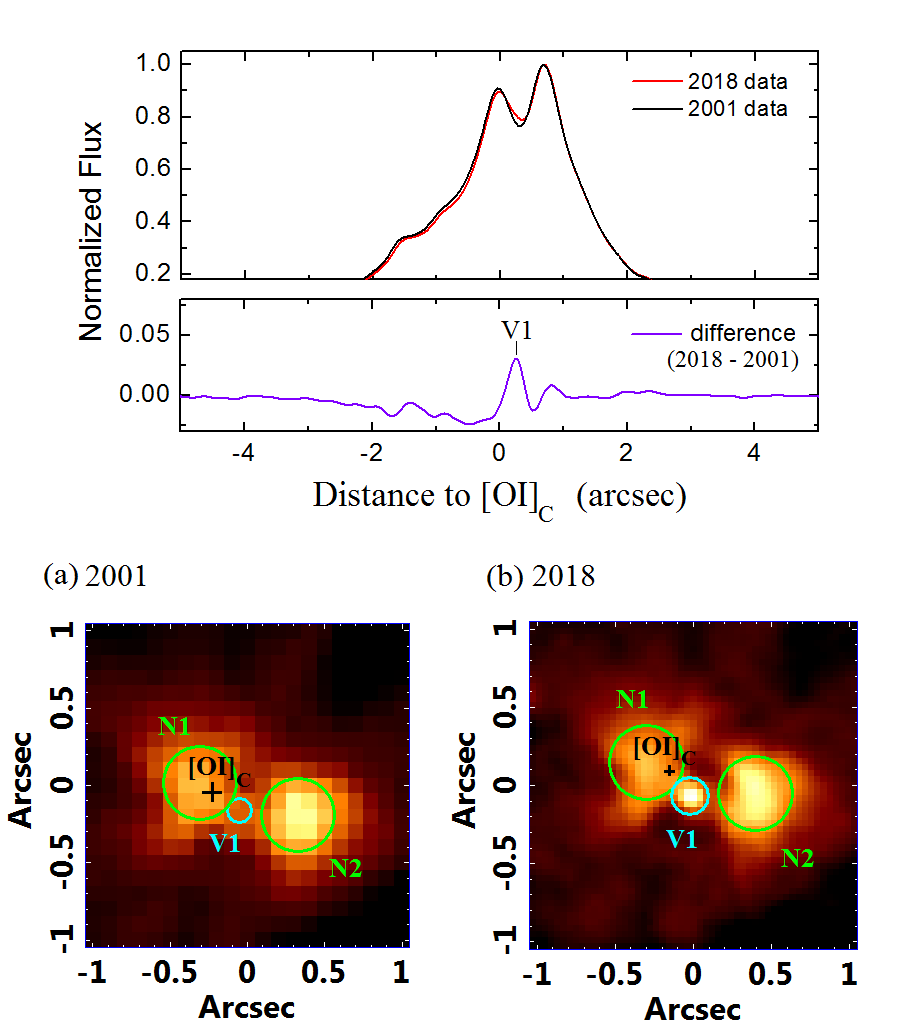}
  \caption{Top: \textit{F 814 W} filter profile extracted along the PA = 255$^{\circ} \pm 5^{\circ}$ (the orientation that passes through [O \textsc{i}]$_{C}$ and N2) from the \textit{HST} images observed on 2001 (WFPC2) and 2018 (WFPC3). The images were normalized considering the flux in N2. Also the image from WFPC3 was convolved with an estimate of the PSF of the image from WFPC2 for a better comparison. The difference between the 2018 data and the 2001 profiles is represented by the purple curve with the indication of the source of variation V1. We noticed a source between N1 and N2 ($\sim$ 0.24 arcsec from [O \textsc{i}]$_{C}$) that has higher luminosity than before. We named this source V1. Bottom: images of \textit{F 814 W} \textit{HST} filter of the inner centre of NGC 613 (a) taken in 2001 and (b) taken in 2018. Both images show indication of regions N1 and N2, [O \textsc{i}]$_{C}$ position and V1. The diameter of the green circles was determined based on the size of N2 in the 2018 image and the diameter of the cyan circle was taken as the size of V1 in the 2018 image. The cross indicating [O \textsc{i}]$_{C}$ corresponds to 3$\sigma$ of the GMOS spaxel.\label{HSTsubtracoes}}
\end{center}
\end{figure}

\subsection{Optical and NIR gas data cubes}\label{starlight_def}

In order to study the gas emission in the data cubes, we applied a spectral synthesis using the \textsc{starlight} software \citep{starlight}, with a stellar population spectral base created using the Medium-resolution Isaac Newton Telescope Library of Empirical Spectra (MILES, \citealt{blazquez}) in the optical spectral range (GMOS and SIFS), to trace the stellar continuum and remove it. The spectral synthesis is applied to each spectrum of the data cube and consists of a linear combination of the base spectra to obtain the observed spectrum. First, we mask all the emission lines of the treated data cube and then we apply the spectral synthesis. After that, we create a synthetic stellar data cube containing only the synthetic spectra obtained in the process. Then, subtract this data cube from the original treated one, obtaining a data cube with mainly gas emission that we call gas data cube.

In order to obtain the SINFONI gas data cube, we performed the spectral synthesis using the Penalized Pixel Fitting (pPXF, \citealt{cappellari}) method. The spectral base covers only the \textit{K} band (\citealt{starlightinfrared}). pPXF is a spectral synthesis that uses the spectra from the base convolved with Gauss--Hermite functions. Then, in the same way as in the optical, we create a synthetic stellar data cube and subtract it from the original one, obtaining a gas data cube, in this case, only in the \textit{K} band.

\subsection{\textit{Chandra} data}\label{chandradata}

The data from \textit{Chandra} space telescope were obtained from the public archive. The observations were taken on 2014 August 21 (PI: Garmire, G., program: 15610062), with a 14.1 h exposure, using the ACIS instrument. From the information of the images headers, referring to the positions and energy of the detected photons, we created a data cube. 

The data cube was spatially re-sampled. The original size of the spaxels was 0.492 arcsec and, after the re-sampling, was 0.246 arcsec. After that, a spatial filtering using the Butterworth method was applied and the data cube was spatially resized, in order to make superpositions (or more direct comparisons) between this data cube and the others analysed in this work. To do this, we took a reference point in this data cube and its values of RA and Dec. and compared to the other data cubes and also the \textit{HST} images, then we centred the images at the same point. 

We used the data of Mrk 202 to compare PSFs, as shown in section \ref{secraiosx}. These data were obtained with the same instrument, ACIS, and the same process of creation, re-sampling and resizing used in the data cube of NGC 613 was performed. The data were taken on 2003 March 16 (PI: Predehl, P., program: 04700038) with a 2.2 h exposure.

\subsection{ALMA data}\label{ALMAdata}

The data from ALMA were obtained from its public archive. The observations were part of the program 2015.1.00404.S (PI: Combes, F.) and these data were already been published in \citet{Combess} and \citet{audibert} works. We used here only the data cube of CO(3-2) emission line, whose beam size was 0.43 arcsec $\times$ 0.37 arcsec and resolution was 0.14 arcsec. Since we just needed the image of the cube collapsed along the spectral axis and the data has a good quality, we did not perform any treatment in this data cube. The spaxel size of the images of this data cube is 0.07 arcsec. 

In the previous works these data were used to determine the geometry of the dust torus of this nucleus \citep{Combess} and the identification of the nuclear spiral by \citep{audibert}, they also compared this data with SINFONI data. Here, we essentially compare these data with the \textit{HST} images, reinforcing a connection between the bar and the nuclear spiral.

\section{Emitting regions}\label{emissao}

The central region of NGC 613 is a complex environment, since it has emissions of many sources and natures. The emitting regions were studied according to the instruments, depending on the spatial resolution and FOV. In other words, the inner central region was mainly studied using the GMOS data cube and \textit{HST} images, whereas the circumnuclear region was better studied using SIFS and SINFONI, \textit{Chandra}, and ALMA data cubes. The gas emission, as said previously (see section \ref{starlight_def}), was analysed using the gas data cubes of GMOS, SIFS and SINFONI.

\subsection{Nuclear emission -- matching the images of distinct instruments}\label{matching}

The integrated GMOS data cube shows an extended emission with an elongated morphology, consistent with the presence of a double structure (see Fig.~\ref{gmoscolapsed}). This same pattern is present both in the emission-line images and in the continuum image of this data cube. Since we did not detect any featureless continuum in this object in the GMOS data cube (see Paper II), we can say that the continuum emission from this double structure is essentially stellar.

The \textit{HST} images also present a double structure that is the same one observed in the GMOS data cube, since it has the same length ($1.65 \pm 0.05$ arcsec) and orientation (PA = 255$^{\circ} \pm 5^{\circ}$), as shown in Fig. \ref{GMOS_HST}. It is worth mentioning that the uncertainties given here and throughout the text are the relative uncertainties, based on the spatial sampling of the images. The absolute uncertainties correspond to the uncertainties of the pointing of the instruments. The emission of [O \textsc{i}]$\lambda$6300 looks point-like in the GMOS data cube and we define the position of the peak of this emission as [O \textsc{i}]$_{C}$, whose coordinates are $X_C= -0.35 \pm 0.03$ arcsec and $Y_C= -0.20 \pm 0.03$ arcsec in the GMOS FOV. When we superpose the GMOS and the \textit{HST} data (Fig. \ref{GMOS_HST}a), we see that the [O \textsc{i}]$\lambda$6300 emission nearly coincides with one of the components of the double stellar structure (the one at NE). We call this component of the double stellar structure as N1. N2 was defined as the brightest stellar nucleus observed in the \textit{HST} image (see Fig. \ref{GMOS_HST}d). In the GMOS data cube, the centre of N2, with coordinates $X_C= 0.1 \pm 0.05$ arcsec and $Y_C= 0.15 \pm 0.05$ arcsec, was defined as the centre of the extended emission, northwest of the [O \textsc{i}]$\lambda$6300 emission peak (obeying the orientation observed in the \textit{HST} images), see Fig. \ref{GMOS_HST}(a).

The centre of N2, visible both in the \textit{HST} and GMOS images, was used as reference to match the images of the two telescopes. The \textit{HST}-based coordinates for the centre of N1 and N2 and the position of [O \textsc{i}]$_{C}$ are given in Table \ref{table_posicoesN1N2}. We used the position of [O \textsc{i}]$_{C}$ to superpose the images of \textit{HST} and SIFS FOV (Fig.\ref{hst_tot}). The SIFS data cube does not have enough spatial resolution to allow the visualisation of the extended emission that we see in the GMOS data cube. Because of that, we were only able to determine the position of [O \textsc{i}]$_{C}$ and, therefore, we could not study region N2 by analysing this data cube.

\begin{table}
\centering
\caption{Right ascension (RA) and declination (Dec.) of [O \textsc{i}]$_{C}$, and the centres of N1 and N2. These coordinates were derived from the \textit{HST} images, as pointed in Fig. \ref{GMOS_HST}d. The relative uncertainties of the values are $\sim 0.02$ arcsec. The absolute uncertainties correspond to the uncertainty of the pointing of the \textit{HST}.} \label{table_posicoesN1N2}
\begin{tabular}{ccc}
\hline
   & RA             & Dec.                                                                                         \\ \hline
[O \textsc{i}]$_{C}$      & 1h 34m 18.172s & --29d 25m 5.885s \\
N1      & 1h 34m 18.201s & --29d 25m 5.74s               \\
N2      & 1h 34m 18.139s & --29d 25m 5.99s  \\ \hline
\end{tabular}
\end{table}

\begin{figure}
\begin{center}
   \includegraphics[scale=0.45]{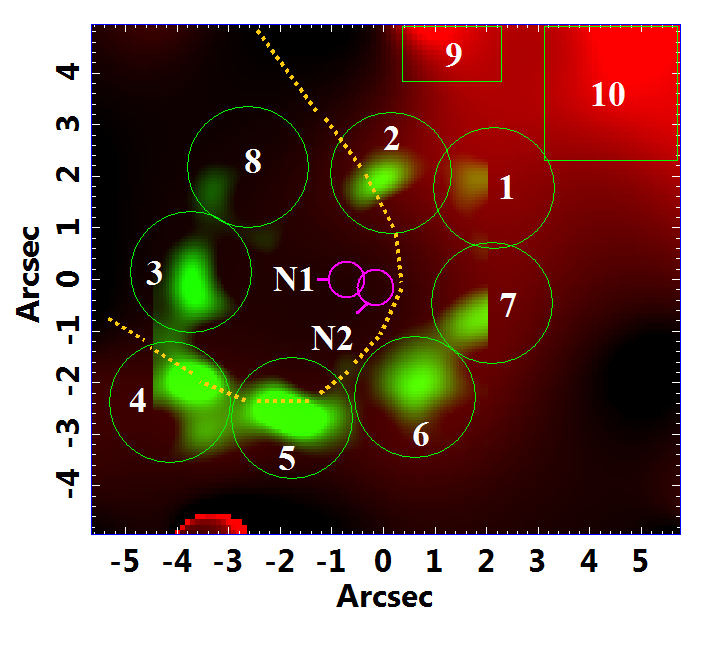}
   \caption{RG composite image, with red representing the image of the H$\alpha$/[N \textsc{ii}]$\lambda$6584 ratio, taken from the SIFS data cube (indicating the areas where the H$\alpha$ emission is higher or more relevant than the [N \textsc{ii}]$\lambda$6584 emission), and the green corresponding to the Br$\gamma$ emission image, from the SINFONI data cube, which delineates the star formation ring. We added a dashed yellow line representing approximately the [O \textsc{iii}]$\lambda$5007 emission (the ionization cone edges). From the Br$\gamma$ image, we identified eight emitting regions (named from 1 to 8). In addition to these, we detected regions 9 and 10, from the image of the SIFS data cube, as they are not part of the SINFONI FOV and neither of the ring. We added circles to the image representing the positions of N1 and N2, whose sizes are the same of the PSF from the GMOS data cube. All the circles represent the extraction area of the spectrum of each region, in order to calculate their emission-line ratios. For N1 and N2, we used the GMOS gas data cube to extract their spectra, because the spatial resolution was better and high enough to separate those emissions. Since the spectra of regions 1 to 10 were extracted from the SIFS gas data cube (because the GMOS FOV does not contain those regions, but only the inner edges of some of them), we used the PSF of this data cube for the extraction. Note that the number that indicates each regions follows the same orientation as \citet{falcon613}, however Region 8 is not the same one the authors named in their article. \label{figextracaoespectro}}
\end{center}
\end{figure}

In the case of the SINFONI data cube, we noted that the Br$\gamma$ inverted image, in flux units, shows a lack or low emission, or even Br$\gamma$ in absorption, in the inner centre of NCG 613. The Br$\gamma$ in absorption is the most likely feature and it might be related to the presence of young stellar populations in regions N1 and N2 (see Paper II), since those stars present deeper hydrogen absorption lines than the older ones. The pattern of the region of Br$\gamma$ in absorption located at the centre is similar to the extended emission that we see in the GMOS data cube (see Fig. \ref{SINFONIN1N2}). When we superpose the image of the point-like source of the SINFONI data cube (H$_2\lambda$21218 in the centre) to the image of the only point-like source of the GMOS data cube ([O \textsc{i}]$\lambda$6300), we find a resemblance between them (compare Fig. \ref{SINFONIN1N2} and Fig. \ref{GMOS_HST}a), with [O \textsc{i}]$_{C}$ slightly displaced towards the centre and with the N2 position compatible with the other end of this extended morphology. From that and also from the fact that the superposition of SINFONI and SIFS (after assuming the positions for [O \textsc{i}]$_{C}$ and for the centre N2) is correct, as it delineates successfully the circumnuclear ring as observed in the SIFS data cube (see Fig. \ref{figextracaoespectro}), we concluded that the position of [O \textsc{i}]$_{C}$ is, within the uncertainties, the centre of the H$_2\lambda$21218 emission. After we made the superposition we noted that, the values of RA and Dec. of N2 and [O \textsc{i}]$_{C}$ are different from the ones we obtained in the optical (Table \ref{table_posicoesN1N2}), indicating that the values of RA and Dec. from the SINFONI data cube are not reliable.

 Fig. \ref{GMOS_HST}(a) shows the result of the GMOS data cube collapsed along its spectral axis, the positions of [O \textsc{i}]$_{C}$ and N2 (discussed previously) and the white contours that represent the emission of the [O \textsc{i}]$\lambda$6300 line. In addition to the fact that there is no [O \textsc{i}]$\lambda$6300 emission coming from N2, we see that this region has blue emission (from the image of the \textit{HST} filter \textit{450 W}), whereas such emission is displaced to north-east of [O \textsc{i}]$_{C}$, but still a significant part of it is inside the region that delineates N1 (see Fig.\ref{GMOS_HST}b). At longer wavelengths (\textit{HST} filter \textit{814 W}), we see clearly N1 and N2 (Fig. \ref{GMOS_HST}c). In this case, N1 seems to have stronger emission in red than N2 (better seen in panel d). We see in Fig. \ref{GMOS_HST}(d) that N1 is in a region predominantly red, possibly highly affected by dust, that may cause the displacement of the blue emission we see in Fig. \ref{GMOS_HST}(b), which might have been coming from the position [O \textsc{i}]$_{C}$ (represented by the cross in the image).

The \textit{Chandra} data cube matched the data by the RA and Dec. In section \ref{secraiosx}, we see that the position of [O \textsc{i}]$_{C}$, N1, and N2 centres are compatible with the centre of the hard X-ray emission, since these data do not have enough spatial resolution to separate those sources. So we took the position of [O \textsc{i}]$_{C}$ as reference to the superposition that was made based on the RA and Dec.  

In order to compare the molecular gas emission obtained with the ALMA data cube of CO(3-2), we made a superposition with the molecular emission observed with H$_2\lambda$21218 from SINFONI data cube as shown in Fig. \ref{ANEL_NIR_ALMA}. This superposition was made to match as much as possible the structure of the circumnuclear ring in both images and also was already done by \citet{audibert}. Here we do this superposition to locate  [O \textsc{i}]$_{C}$ and N2 in this context. For discussion of Fig. \ref{ANEL_NIR_ALMA}, see section \ref{circumnuclear_ring}.

\begin{table*}
\centering
\caption{Emission-line ratios of the regions detected in the nucleus of NGC 613, calculated from the integrated flux of the emission-lines of the spectra extracted from each region (see Figs. \ref{espectrosN1N2}, \ref{espectros12345} and \ref{espectros678910} and appendix \ref{espectros_analise}).}\label{razaodelinhastable}

\begin{tabular}{ccccccc}
\hline
Regions & {[}O \textsc{iii}{]}/H$\beta$ & {[}N \textsc{ii}{]}/H$\alpha$ & {[}O \textsc{i}{]}/H$\alpha$ & {([}S \textsc{ii}{]}$\lambda$6716 +$\lambda$6731)/H$\alpha$ & {[}S \textsc{ii}{]}$\lambda$6716/{[}S \textsc{ii}{]}$\lambda$6731 \\ \hline
N1       & 0.72 $\pm$ 0.07                               & 1.87$\pm$ 0.03                                & 0.30 $\pm$ 0.03                              & 1.03 $\pm$ 0.07                                                    & 1.199 $\pm$ 0.017                                                                                 \\
N2     & 1.55 $\pm$ 0.23                               & 1.19 $\pm$ 0.06                               & 0.184 $\pm$ 0.023                            & 0.69 $\pm$ 0.15                                                    & 1.09 $\pm$ 0.05                                                                                   \\
1      & 0.24 $\pm$ 0.05                               & 0.38 $\pm$ 0.03                               & 0.031 $\pm$ 0.004                            & 0.194 $\pm$ 0.012                                                  & 1.36 $\pm$ 0.15                                                                                  \\
2       & 1.04 $\pm$ 0.06                               & 0.73 $\pm$ 0.04                               & 0.075 $\pm$ 0.009                            & 0.35 $\pm$ 0.05                                                    & 1.33 $\pm$ 0.04                                                                                    \\
3        & 1.13 $\pm$ 0.09                               & 0.888 $\pm$ 0.021                             & 0.089 $\pm$ 0.010                            & 0.46 $\pm$ 0.07                                                    & 1.34 $\pm$ 0.04                                                                                   \\
4       & 0.240 $\pm$ 0.020                             & 0.417 $\pm$ 0.016                             & 0.025 $\pm$ 0.006                            & 0.225 $\pm$ 0.013                                                  & 1.25 $\pm$ 0.12                                                                                   \\
5        & 0.136 $\pm$ 0.010                             & 0.394 $\pm$ 0.005                             & 0.0124 $\pm$ 0.0016                          & 0.172 $\pm$ 0.004                                                  & 1.18 $\pm$ 0.05                                                                                   \\
6       & 0.166 $\pm$ 0.019                             & 0.371 $\pm$ 0.008                             & 0.024 $\pm$ 0.004                            & 0.185 $\pm$ 0.006                                                  & 1.12 $\pm$ 0.07                                                                                    \\
7        & 0.37 $\pm$ 0.06                               & 0.525 $\pm$ 0.022                             & 0.092 $\pm$ 0.010                            & 0.29 $\pm$ 0.07                                                    & 1.23 $\pm$ 0.06                                                                                   \\
8      & 1.91 $\pm$ 0.18                               & 1.47 $\pm$ 0.06                               & 0.15 $\pm$ 0.03                              & 0.75 $\pm$ 0.13                                                    & 1.44 $\pm$ 0.05                                                                                   \\
9        & 0.3 $\pm$ 0.3                                 & 0.42 $\pm$ 0.08                               &                                              &                                                                    &                                                                                                   \\
10      & 0.19 $\pm$ 0.11                               & 0.25 $\pm$ 0.03                               &                                              & 0.143 $\pm$ 0.013                                                  & 1.25 $\pm$ 0.23                                                                                  \\ \hline
\end{tabular}
\end{table*}

\begin{table}
\centering
\caption{FWHM and luminosity of the H$\alpha$ line determined for each region (N1 and N2 from the GMOS data cube spectra and the other regions from the SIFS data cube spectra). The spectra were corrected of reddening using H$\alpha$/H$\beta$ ratio.}\label{tabelacoisas}
\begin{tabular}{ccccc}
\hline
Regions & \begin{tabular}[c]{@{}c@{}}FWHM of H$\alpha$\\ (km s$^{-1}$)\end{tabular} & \begin{tabular}[c]{@{}c@{}}H$\alpha$ luminosity\\  (10$^{39}$ erg s$^{-1}$)  \end{tabular} \\ \hline
N1      & 592 $\pm$ 8                                                                & 3.95 $\pm$ 0.06                                                                            \\
N2      & 258 $\pm$ 7                                                                & 9.2 $\pm$ 0.3                                                                              \\
1       & 138 $\pm$ 5                                                                & 40.0 $\pm$ 1.2                                                                             \\
2       & 143 $\pm$ 5                                                                & 32.8 $\pm$ 1.0                                                                             \\
3       & 172 $\pm$ 5                                                                & 26.0 $\pm$ 0.4                                                                             \\
4       & 106 $\pm$ 5                                                                & 73 $\pm$ 2                                                                                 \\
5       & 113 $\pm$ 5                                                                & 65.1 $\pm$ 0.4                                                                             \\
6       & 122 $\pm$ 5                                                                & 59.0 $\pm$ 0.3                                                                             \\
7       & 226 $\pm$ 5                                                                & 37.1 $\pm$ 0.7                                                                             \\
8       & 331 $\pm$ 7                                                                & 6.61 $\pm$ 0.21                                                                            \\
9       & 114 $\pm$ 7                                                                & 0.50 $\pm$ 0.04                                                                            \\
10      & 183 $\pm$ 5                                                                & 14.8 $\pm$ 0.5    \\ \hline
\end{tabular}
\end{table}

\subsection{Nuclear variability}

When we compare the collapsed GMOS data cube images and the \textit{HST} RGB composition we see a difference between the emissions of N1 and N2: in the \textit{HST} images, N2 is the brightest source and, in the GMOS data cube image, N1 is brighter than N2. In order to make a more direct comparison of this feature, the GMOS data cube was integrated along the spectral axis, taking into account the transmission curve of the \textit{HST} filter \textit{F 606 W}. Besides that, the image of the \textit{HST} filter \textit{F 606 W} was re-sampled to have pixels with the same sizes of the spaxels of the GMOS data cube (0.05 arcsec $\times$ 0.05 arcsec), filtered with the Butterworth method (since the re-sampling generates high frequency noise) and, then, convolved with the PSF of the GMOS data cube (0.68 arcsec). Both images were normalized in N2 (the emission peak in N2 became equal to one), since the flux units are different in these two instruments. From this, we made a direct comparison between N1 and N2, as shown in Fig. \ref{GMOSMENOSHST}. The difference between the final image from the GMOS data cube and the final image from the \textit{HST} filter \textit{F 606 W} (Fig. \ref{GMOSMENOSHST}c) shows a source whose centre is compatible with the position of [O \textsc{i}]$_{C}$, within the uncertainties (the distance between [O \textsc{i}]$_{C}$ and this source is 0.11 arcsec). Such a source, which we call V2, was not detected before. The brightness variation of this source happened in the interval of about 14 yr (between 2001 August and 2015 January, when the observations of \textit{HST} and GMOS were taken, respectively). Such a variation could be an evidence of a supernova. However, one can notice a vertical pattern in Fig. \ref{GMOSMENOSHST}(c) that is similar to the instrumental fingerprint signature in the GMOS data cube images (see \citealt{gmostreat}). We cannot discard that this image may have some instrumental fingerprint remnants, still present in the GMOS data cube after treatment, that generated this pattern. Nevertheless, we know that the instrumental fingerprint cannot generate emissions that are approximately point-like as the one that we are seeing. 
 
 Another way to study variability was possible due to the availability of recent images from the \textit{HST}. By comparing images obtained with the same filter (\textit{F 814 W}) in 2001 and 2018, we aimed to evaluate in more detail possible flux variations of the sources during this period. The 2001 image was obtained with WFPC2, while the 2018 image was obtained with WFPC3, which has a higher spatial resolution. In order to perform a comparison, first of all, we convolved the 2018 image with an estimate of the PSF of the 2001 image. After that, we extracted, from each image, a brightness profile from a rectangular region along a PA= 255$^{\circ}$ and with a width of 0.75 arcsec. The two extracted profiles were normalized in a way that the flux peak (in N2) in both profiles was equal to 1. The top panel of Fig. \ref{HSTsubtracoes} shows the two extracted profiles, together with the curve corresponding to the difference between them. We notice a flux increase, from 2001 to 2018, in a source between N1 and N2, whose position is not compatible, even at the 3$\sigma$ level, with the position of [O \textsc{i}]$_{C}$. We named this variable source as V1. The distance between V1 and [O \textsc{i}]$_{C}$ is $0.24 \pm 0.04$ arcsec. This variation is better seen by comparing the images of the bottom panel in Fig. \ref{HSTsubtracoes}. If the real position of the AGN is V1, we are seeing that this AGN is variable. If not, we are seeing another indication of supernova or even variations of dust extinction in the line of sight. Similar to V2, the presence of V1 has not been previously reported in the literature.

\begin{figure}
\begin{center}
   \includegraphics[scale=0.41]{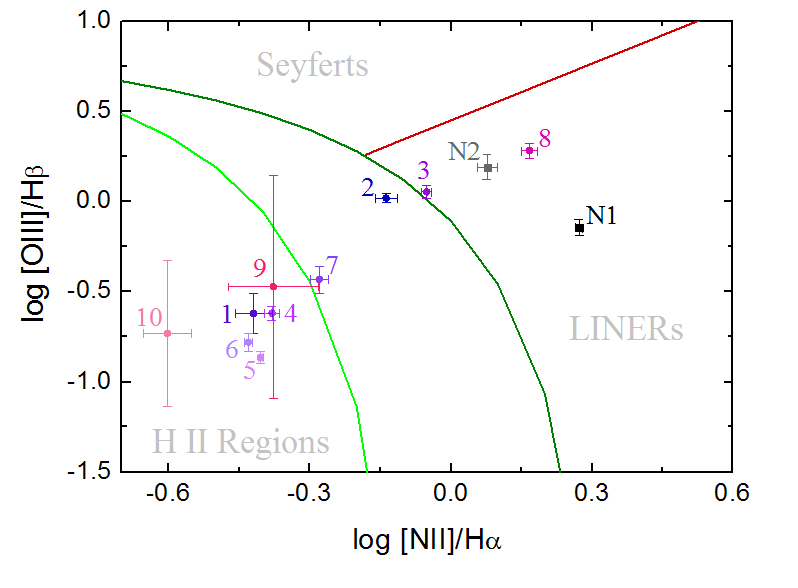}
  \caption{Diagnostic diagram obtained from the emission-line ratios on table \ref{razaodelinhastable}. The names of each studied region are indicated near their respective point. The dark green line represents the theoretical limit for the starburst emission obtained by \citet{kewley1}. The separation between H \textsc{ii} regions and AGNs determined by \citet{kauffmann} is indicated by the light green curve. The red line represents the separation between the Seyferts and LINERs emission determined by \citet{schawinski}. The other two diagnostic diagrams,log [O \textsc{iii}]/H$\beta$ $\times$ log ([S \textsc{ii}]$\lambda$6716 + $\lambda$6731)/H$\alpha$ and log [O \textsc{iii}]/H$\beta$ $\times$ log [O \textsc{i}]/H$\alpha$ represents essentially the same results, and therefore are not displayed in this work.}\label{diagdiag}
\end{center}
\end{figure}

\begin{figure}
\begin{center}
   \includegraphics[scale=0.4]{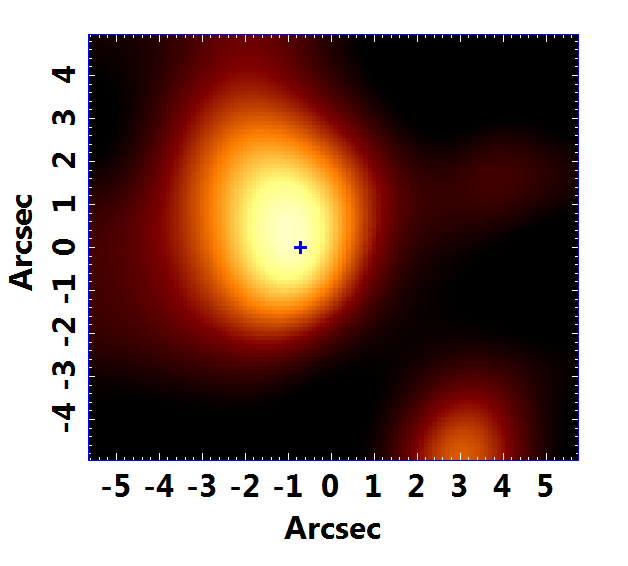}
   \caption{[O \textsc{iii}]$\lambda$5007 image obtained from the SIFS data cube. The blue cross represents the position of [O \textsc{i}]$_{C}$ and its size the uncertainty of 3$\sigma$. \label{OIIIimagem}}
\end{center}
\end{figure}

\subsection{H II regions}

It is known, in the literature, that NGC 613 nucleus has a circumnuclear ring with many star-forming regions (\citealt{boker}; \citealt{falconIAU,falcon613}). The FOV of the GMOS data cube (3 arcsec $\times$ 5 arcsec) is restricted only to the inner centre of the galaxy and its size does not cover the region of the circumnuclear ring, but only part of its inner edges. For this reason we analysed the circumnuclear ring with the SIFS data cube, which has a larger FOV than the GMOS data cube. So, to study these regions, we used the SIFS and SINFONI data cubes (the latter was the one used in the literature to characterize the ring in the NIR). 

The Br$\gamma$ emission image of the SINFONI data cube clearly shows eight star-forming regions that compose the ring (see Fig. \ref{figextracaoespectro} in green). The ring can be observed in the SIFS data cube, however this data cube does not have enough spatial resolution to separate each region. On the other hand, as we can see in Fig. \ref{figextracaoespectro} (in red), we detected, besides the ring, two regions (named 9 and 10) that seem to be connected to the ring. \citet{falcon613} detected seven H \textsc{ii} regions, that are regions 1 to 7 in Fig. \ref{figextracaoespectro}. However, we are considering another region in the ring (Region 8) and their region 8 is actually a region inside the ring, but not part of it, that they use for comparison.

\begin{figure}
\begin{center}
   \includegraphics[scale=0.3]{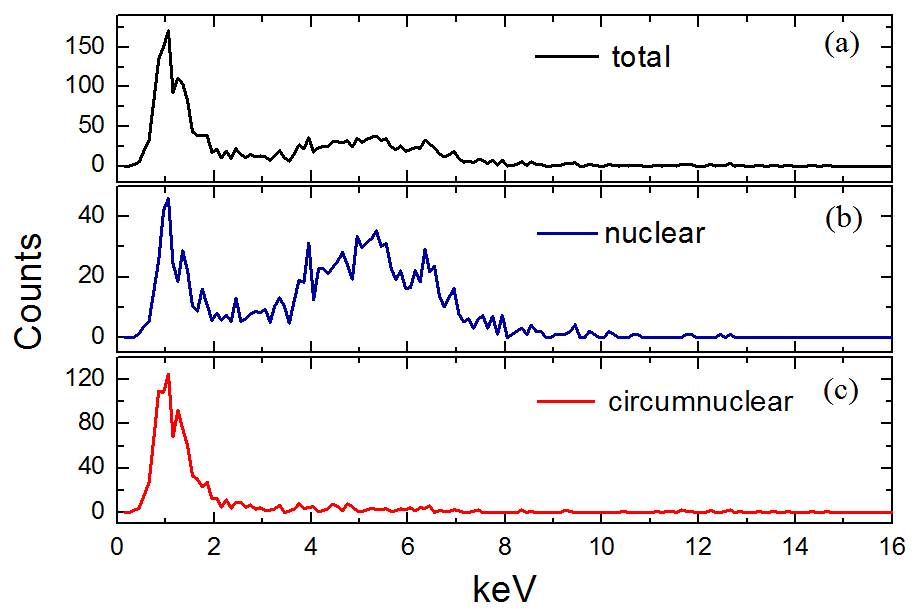}
   \caption{(a) Total spectrum of the \textit{Chandra} data cube of NGC 613, (b) spectrum of the circular region whose centre is indicated by the blue cross in Fig. \ref{raioxduros} with the extraction radius of $\sim$ 1.7 arcsec and (c) the total spectrum of the area of $\sim$ 20 arcsec $\times$ 20 arcsec (same area of the FOV of the Fig. \ref{raioxduros}) minus the one from the nuclear region, representing the circumnuclear spectrum. \label{espectrochandra}}
\end{center}
\end{figure}

\begin{figure*}
\begin{center}
   \includegraphics[scale=0.4]{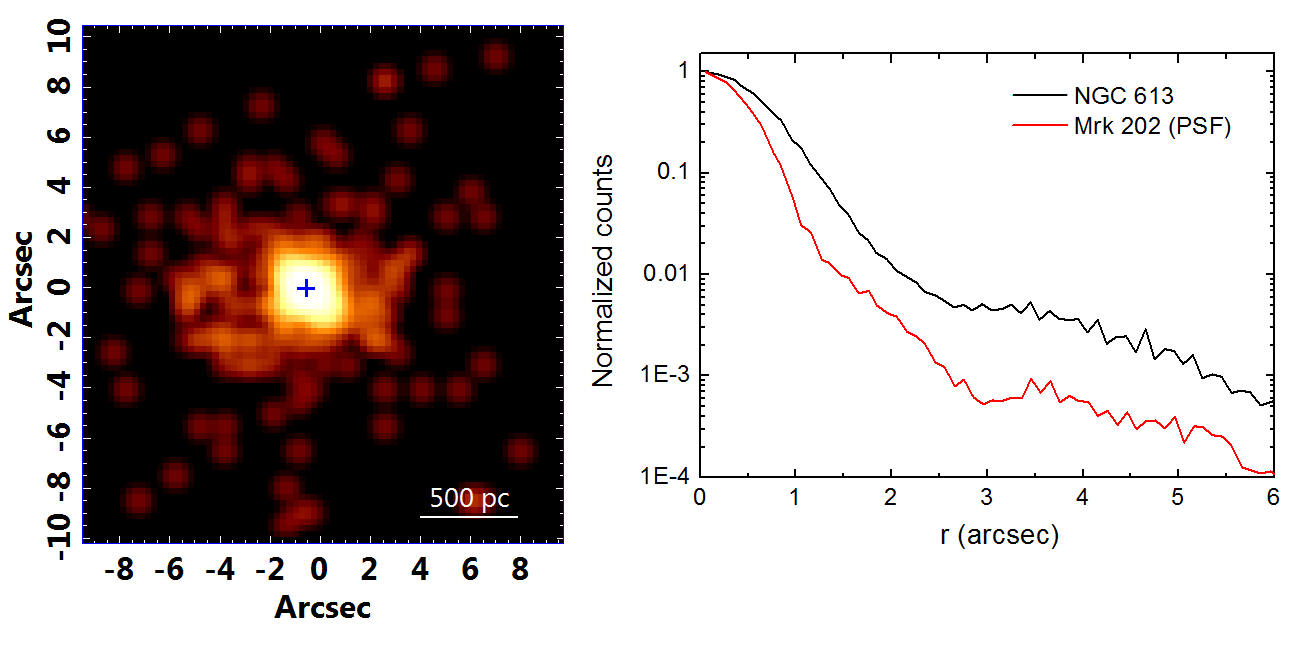}
   \caption{Hard X-ray image, 2 to 10 keV, of the \textit{Chandra} data cube of NCG 613 and radial profiles of the images of hard X-ray from NGC 613 and Mrk 202 (a Seyfert 1 galaxy observed with the same instrument, used here as an estimate of the PSF of the data, since it has only a central emission). The blue cross indicates the position of the peak of the emission and its size the uncertainty of 3$\sigma$. \label{raioxduros}}
\end{center}
\end{figure*}

\subsection{Optical emission-line ratios }\label{sec_razaodelinhas}

From the optical gas data cubes, in this case GMOS and SIFS, it is possible to use the emission-line ratios to characterize the nature of the emission from the regions we detected. Fig. \ref{figextracaoespectro} shows the sizes and positions of the areas that we delineated to extract the spectrum of each region. The spectra of these regions were extracted from this data cube and the diameter of the extraction area was equal to the FWHM of the PSF of these data (see Fig. \ref{figextracaoespectro}, magenta circles). For the regions named 1 to 10, we had to make a superposition of the Br$\gamma$ emission image (which clearly shows regions 1 to 8 in the ring, in green) and of the H$\alpha$/[N \textsc{ii}]$\lambda$6584 image from the SIFS data cube, in red (this image shows where H$\alpha$ emission is more relevant and therefore the presence of the H \textsc{ii} regions), since we had to know the positions of those regions in this data cube, in order to extract the spectra to calculate the emission-line ratios. As the spectra were extracted from the SIFS data cube, the diameter of the extraction areas of those 10 regions was equal to the FWHM of the PSF of these data (see green circles in Fig. \ref{figextracaoespectro}), with exception of regions 9 and 10 that had their areas and positions estimated directly from the image of H$\alpha$/[N \textsc{ii}]$\lambda$6584 and, as they are at the edges of the FOV, we could not extract circular areas (see green rectangles in  Fig. \ref{figextracaoespectro}).  

The extracted spectra are presented in Appendix  \ref{espectros_analise}. As one can see, many spectra presented blended emission lines, without indications of broad components of the H$\alpha$ and H$\beta$ lines. In order to calculate the emission-line ratios, we decomposed those blended lines in Gaussians functions. The decomposition method and the flux and decomposition uncertainties are also discussed in appendix \ref{espectros_analise}. Table \ref{razaodelinhastable} contains the emission-line ratios with their respective uncertainties of the 12 regions and Fig. \ref{diagdiag} shows the diagnostic diagram of [O \textsc{iii}]$\lambda$5007/H$\beta$ $\times$ [N \textsc{ii}]$\lambda$6584/H$\alpha$.  

We can see that, excepting regions N1, N2, 2, 3, and 8, the emission of all other ones is compatible with the emission from H \textsc{ii} regions. N1 and N2 have emission-line ratios compatible with the ones of LINERs. Based on the [O \textsc{iii}]$\lambda$5007/H$\beta$ ratio, N2 seems to have a higher ionization degree than N1, although the values of this ratio for N1 and N2 are compatible, at the 3$\sigma$ level. Regions 2 and 3 have emission-line ratios compatible with the ones of transition objects. The yellow contour of Fig. \ref{figextracaoespectro} corresponds to the edges of the emission of [O \textsc{iii}]$\lambda$5007 (an emission line of relatively high ionization degree) that, by its morphology, characterises an ionization cone (see Fig. \ref{OIIIimagem}). The fact that regions 2, 3, and 8 have such ionization degrees (based on the [O \textsc{iii}]$\lambda$5007/H$\beta$ ratio) and emission-line ratios compatible with the ones of AGNs indicates that the emission from those regions is highly contaminated by the ionization cone (being either the result of an overlap of emissions or of an ionization of these regions directly by the AGN). Region 8 is also very close to the position of the radio jet and outflow and this resulting ionization degree might be due to a contamination from the radio jet and outflows coming from the AGN (see section \ref{secraiosx}).

We will discuss in section \ref{secraiosx} that the [O \textsc{iii}]$\lambda$5007 emission is co-spatial with the emission of soft X-ray. There is also a source in the FOV edge at south that can be the other side of this ionization cone or a region of the narrow-line region (NLR) that is receiving the ionizing radiation from the AGN with less obscuration. In Fig. 10 from \citet{gadotti2019}, we see that this source is also extended and it has a high ionization degree, but it was not detected in soft X-ray images. That might be due to extinction in the line of sight or to the distance of this region from the AGN. But the absence of the X-ray emission of this region indicates that it is not the position of the AGN.

Table \ref{tabelacoisas} presents the FWHM and luminosity of the H$\alpha$ emission line in the spectra of the observed regions. One can note that N2 is brighter than N1 (regarding H$\alpha$ emission) and that region 4 is the brightest of the ring. N1 has the highest FWHM values, which indicates the presence of outflows of gas. 

\subsection{Absence of featureless continuum emission in the \textit{K} band}\label{nofeatcont}

A map of the $D_{CO}$ coefficient \citep{marmol} obtained from the same SINFONI data analysed in this work is shown in Fig. 5 of \citet{falcon613}. The values of such a coefficient are higher for spectra with a deeper CO absorption band at $2.29$ $\mu$m. The authors note a slight decrease in the values of the $D_{CO}$ coefficient at the nucleus, in comparison to the surroundings. If this object presented a significant emission of a featureless continuum, we would expect shallower absorption CO bands at the nucleus (see \citealt{burtscher}), which would result in lower values of $D_{CO}$ in that region. Considering the fact that the $D_{CO}$ map presented by \citet{falcon613} is quite noisy (due probably to the S/N of SINFONI data cube), it is difficult to confirm whether or not the drop of the values at the nucleus is real. We conclude that the effect of a featureless continuum in these data is, at most, very weak. Therefore, the featureless continuum at the nucleus of NGC 613, in the \textit{K} band, is either too weak (due to an AGN lower state of activity, if it is variable) to be detected or should be highly obscured in the \textit{K} band. See section \ref{double_stellar_agn} for a detailed discussion and a comparison with previous studies.

\section{The X-ray emission}\label{secraiosx}

\begin{figure}
\begin{center}
   \includegraphics[scale=0.45]{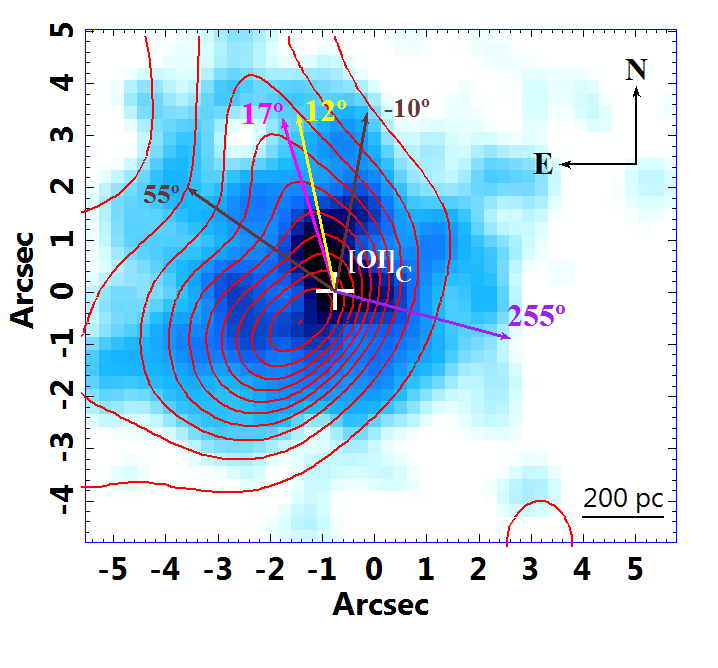}
  \caption{Image from \textit{Chandra} space telescope of the central region of NGC 613 in the soft X-rays, with indication of the orientation N-E and scale of 200 pc. The PA of the radio jet detected by \citet{hummel} is represented by the yellow vector (PA = 12$^{\circ}$), the PA of the outflow observed in the H$\alpha$ channel map with v= 306 km s$^{-1}$ (see Paper II), with PA = 17$^{\circ}$, is represented by the magenta vector and the outflows observed in [O \textsc{iii}]$\lambda$5007 with v= --889 km s$^{-1}$ (PA $\sim$ --10$^{\circ}$) and with v = --710 km s$^{-1}$ (PA $\sim$ 55$^{\circ}$) are represented by the brown vectors. The purple vector with PA = 255$^{\circ}$ connects [O \textsc{i}]$_{C}$ to N2, but [O \textsc{i}]$_{C}$ and the centre of N2 are approximately at the same point, because of the low spatial resolution. [O \textsc{i}]$_{C}$ position was taken as the centre of hard X-rays emission, represented by the white cross and its size is the 3$\sigma$ uncertainty. We also added the red contours that represent the ionization cone observed in the image of the [O \textsc{iii}]$\lambda$5007 line. \label{chandramoles}}
\end{center}
\end{figure}

In order to further investigate the nature of the emission in the centre of NGC 613, we analysed a data cube obtained with the \textit{Chandra} space telescope. We subtracted the spectrum of a circular region centred on the nucleus with a radius of $\sim$ 1.7 arcsec (Fig.\ref{espectrochandra}b) from the spectrum of the total data cube (Fig. \ref{espectrochandra}a), resulting in the spectrum of the circumnuclear region (Fig.\ref{espectrochandra}c). It is clear that the nucleus shows both hard (2--10 keV) and soft (0.5--2 keV) X-ray emission while the circumnuclear region is dominated by soft X-ray emission. By comparing the circumnuclear spectrum with high-resolution data from NGC 1068 \citep{NGC1068} we identified the lines of Ne \textsc{ix} at 0.91 keV, Ne \textsc{x} Ly$_{\alpha}$ at 1.03 keV and Ne \textsc{x} Ly$_{cont}$ at 1.35 keV. 

Fig. \ref{raioxduros} shows that the hard X-ray emission is concentrated, mainly, in the inner centre. In order to see how compact the central emission is, we compared its profile with that of Mrk 202, a Seyfert 1 galaxy \citep{mr202class} observed with the same instrument as NGC 613 (see section \ref{chandradata}). The image of Mrk 202 AGN in hard X-rays shows only a central source and no relevant circumnuclear features. The radial profiles of the hard X-ray emission of those galaxies are clearly distinct (right-hand panel of Fig. \ref{raioxduros}). The FWHM of the hard X-ray emission of NGC 613 is 1.32 arcsec ($\sim$ 170 pc), while the FWHM of Mrk 202 is 0.99 arcsec; the difference is 0.33 arcsec. 

There is also a clear extended emission with an elongated structure that is consistent with being an inclined disc, as seen in radio, [Fe \textsc{ii}]$\lambda$16436, Br$\gamma$ and molecular lines (Fig. \ref{raioxduros}). The best estimates of the radius and inclination, assuming it is a circular disc, are consistent with the disc parameters seen in the NIR and radio (see table \ref{anelparametros}). This detection of the circumnuclear ring in hard X-rays has not been previously reported in the literature. One may wonder what the origin of such circumnuclear hard X-ray emission is. Our interpretation is that it is most likely originated from supernova remnants (SNRs) associated with the young stellar population. The presence of SNRs in the central region of NGC 613 has already been proposed by \citet{boker} and \citet{falcon613}. See section \ref{circumnuclear_ring} for a detailed discussion.

Fig. \ref{chandramoles} shows an image of the soft X-ray (0.5--2 keV) emission (in blue) with the [O \textsc{iii}]$\lambda$5007 emission contours (in red). There is a good coincidence between the soft X-ray emission and the [O \textsc{iii}]$\lambda$5007 emission which is probably representing the ionization cone in the optical; this clearly shows that there is highly ionized gas in the ionization cone. The emission peak of the ionization cone is not coincident with the emission peak in soft and hard X-rays. On the other hand, there is an enhancement of soft X-ray emission towards the PA of the radio jet and the outflows seen in the H$\alpha$ and [O \textsc{iii}]$\lambda$5007 channel maps (see Paper II). Besides that, one can see that there is also an extended emission towards the PA of the vector that connects N1 and N2.

\section{Discussion}\label{secdiscussion}

NGC 613 has a rich nuclear region. The presence of an AGN \citep{veron}, of a star-forming ring ( \citealt{hummel}, \citealt{BOKERIAU,boker}, \citealt{falconIAU,falcon613}, \citealt{miyamoto1,miyamoto2}), of a radio jet (\citealt{hummel2}, \citealt{hummel}, \citealt{miyamoto1,miyamoto2}), of a nuclear spiral \citep{audibert}, of an outflow (detected from the study of the [O \textsc{iii}]$\lambda$5007 emission by \citealt{hummel2}) and of shock waves \citep{daviesbecca} has already been reported in the literature, making this environment interesting to be studied and challenging to be understood. 

In this section we are presenting all the hypotheses that can explain what we observed considering all the information that we gathered. Section \ref{double_stellar_agn} discusses mainly the coexistence of N1 and N2. In Section \ref{circumnuclear_ring} we present all the hypotheses about the circumnuclear ring and molecular gas emission, taking into account previous studies. Finally, Section \ref{cenariogeraldiscuss} summarizes the whole picture of the nuclear scenario that we believe to be observing, considering all the main structures observed in this work and also previous ones.

\subsection{The double stellar nucleus and the AGN}\label{double_stellar_agn}

The \textit{HST} images of NGC 613 show two sources of stellar emission separated by a stream of dust. The projected distance between these two stellar nuclei is 94 $\pm$ 5 pc ($0.74 \pm 0.04$ arcsec), as seen in \textit{HST} images. These two stellar sources could be seen in \textit{HST} images shown in \citet{falcon613}, \citet{Combess} and \citet{audibert}, but were not discussed by those authors. When we observe the image of the GMOS data cube, we see that the most intense stellar emission is extended and its orientation is the same of the one between the double stellar source in the \textit{HST} images (PA = 255$^{\circ} \pm 5^{\circ}$), as shown in Fig. \ref{GMOS_HST}. We named this emission as regions N1 and N2. We also defined the peak of the [O \textsc{i}]$\lambda$6300 emission as [O \textsc{i}]$_{C}$, which is an emission line typically associated with regions of partial ionization normally seen in AGNs. 

\begin{figure}
\begin{center}
   \includegraphics[scale=0.35]{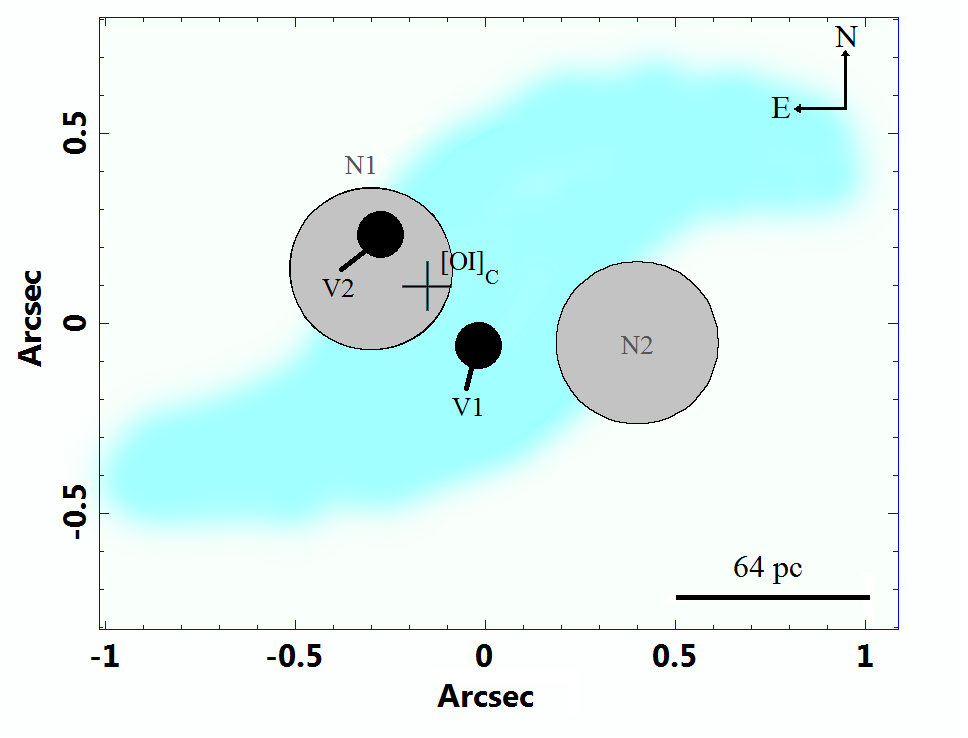}
  \caption{ Scheme of the scenario proposed by this work in the central 2 arcsec. The spiral of molecular gas observed with ALMA is represented by the blue area. The double stellar nucleus is represented by the two grey circles, whose sizes are based on the emission area of N2 in the \textit{HST} images. V1 and V2 are the two variable sources that were detected by comparing \textit{HST} images of filter \textit{F 814 W} in different epochs and of filter \textit{F 606 W} with GMOS data, respectively. Their sizes represent the 3$\sigma$ uncertainty of the position considering the GMOS spaxel size in arcsec. [O \textsc{i}]$_{C}$ is the centre of the emission of [O \textsc{i}]$\lambda$6300 represented by the black cross, whose size is the 3$\sigma$ uncertainty taking into account the spaxels of GMOS. The scale of 0.5 arcsec (64 pc) is indicated in the figure.  \label{scenario}}
\end{center}
\end{figure}

When we compared the \textit{HST} images from 2001 and 2018, we found that there is a source between N1 and N2 that suffered a variation in brightness (Fig. \ref{HSTsubtracoes}).  We call this source, not previously reported in the literature, V1 (see Fig. \ref{scenario}). 

\begin{figure*}
\begin{center}
   \includegraphics[scale=0.4]{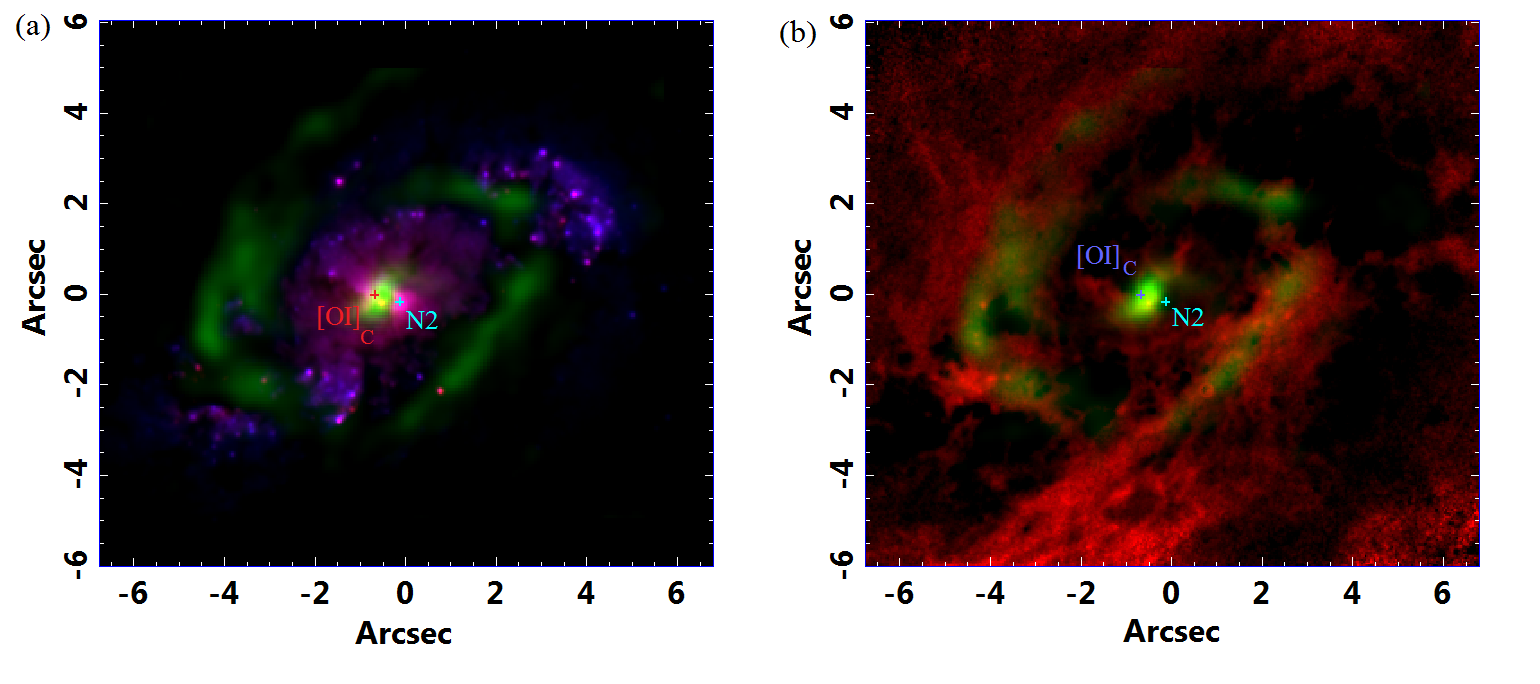}
  \caption{ (a) RGB composition of the \textit{HST} filters \textit{F 814 W} in red, \textit{F 475 W} in blue, and CO(3-2) from ALMA data cube in green. The red and cyan crosses represent the position of [O \textsc{i}]$_{C}$ and the centre of N2, respectively, and its size the uncertainty of 3$\sigma$, considering the size of the pixel of the \textit{HST} images (0.03962 arcsec), since those positions were plotted taking the \textit{HST} images as reference. (b) RB composition of \textit{F 475 W} -- \textit{F 814 W} (\textit{B}--\textit{I} in magnitude scale) in red and CO(3-2) from the ALMA data cube in green. The pink and white crosses represent the position of [O \textsc{i}]$_{C}$ and the centre of N2, respectively, and its size the uncertainty of 3$\sigma$, considering the size of the pixel of the \textit{HST} images. \label{HST_ALMA}}
\end{center}
\end{figure*}

The composition of the \textit{HST} images and the CO(3-2) image from the ALMA data cube (Fig.\ref{HST_ALMA}a) reveals that there is a nuclear spiral that passes between N1 and N2 and the centre of this nuclear spiral is this variable source (V1) that appears strongly in the \textit{F 814 W} filter image from 2018. This nuclear spiral was first detected by \citet{Combess} and \citet{audibert}. V1 suffers obscuration by dust (as we can see in Fig. \ref{hst_tot} and that is compatible also with previous studies that said that NGC 613 AGN is very obscured -- see \citealt{castangia2013}; \citealt{ASMUS2015}) that can be a result of the nuclear spiral that seems to bring gas and dust to the centre.

Our analysis shows that V1 and [O \textsc{i}]$_{C}$ are separated by $0.24 \pm 0.04$ arcsec. The positions of these two sources are not quite compatible, even at the 3$\sigma$ level. One may wonder where the actual position of the AGN really is, in this case. If the emission of [O \textsc{i}]$\lambda6300$ is extended, this shift can be explained by differential dust extinction. However, we also have to acknowledge that, if we think that the original emission of [O \textsc{i}]$\lambda$6300 is coming from V1 (and here assuming that V1 is the AGN), we note that the displacement of such emission is towards the ionization cone. To explain such displacement we know that there is a zone of partial ionization in the walls of the cones where low ionization species are emitted \citep{may}. In addition, some reflection of the nuclear source is also expected in the cones \citep{tiago}. If differential extinction affects more one cone than the other, a small shift of the [O \textsc{i}]$_{C}$ is expected in the direction of the cone with less extinction. The source V1 may be equally reflected but its image was taken with the \textit{F 814 W}  \textit{HST} filter, suffering, thus, less extinction than [O \textsc{i}]$\lambda$6300. So, the main hypothesis here is that AGN is the central variable source (V1) and also the centre of the nuclear spiral. This scenario is in disagreement with Fig. 1 of \citet{audibert}, which suggests that the AGN is located at the position of N2.

Besides the variation of V1, another evidence for the AGN variability is the fact that the $D_{CO}$ map obtained by \citet{falcon613} revealed almost no variations towards the nucleus, as expected in the case of obfuscation by a featureless continuum from an AGN \citep{burtscher}. This is somewhat surprising as the \textit{K} band is less affected by interstellar extinction than the optical. When we look at the image of \textit{F 814 W} filter of \textit{HST} observed in 2018 we see that there is a strong emission in V1 and the possible explanation, then, is that the AGN, in the epoch when the SINFONI data were observed (2005), was probably in a lower state of activity than in 2018.

N1 and N2 show emission-line ratios compatible with those of LINERs (Fig. \ref{diagdiag}). \citet{daviesbecca} obtained optical emission-line ratios for the central region of NGC 613, but using 3D spectroscopy data with a lower spatial resolution than that of the data used in this work. The values of the [O \textsc{iii}]$\lambda$5007]/H$\beta$ ratio, which is used in this work as a measure of the ionization degree, in the spectra of N1 and N2 are compatible, at the 3$\sigma$ level, mainly due of the high uncertainty of this ratio in the spectrum of N2. Even so, we must consider the possibility that N2 may indeed have a higher ionization degree than N1.

By comparing the \textit{HST} images with the GMOS data cube images (Fig. \ref{GMOSMENOSHST}), we see a variable stellar source, not previously reported in the literature, very close to [O \textsc{i}]$_{C}$, inside the area corresponding to N1 (V2 in Fig. \ref{scenario}). The time interval of this variability was 14 yr and it might be due to a supernova, since N1 is a stellar nucleus with the presence of young stellar populations (see Paper II).

As we see in Fig.\ref{raioxduros}, the hard X-ray radial profile shows a width that is larger than the PSF of the instrument, represented here by the radial profile of Mrk 202, a type 1 AGN observed with the same instrument. Since there is no detectable second AGN, it is most likely that this additional emission might be associated with the radio jet in the inner region or that these photons are the result of scattering of the central AGN emission in the circumnuclear region.

The AGN ionization cone is well represented by the [O \textsc{iii}]$\lambda$5007 emission. \citet{daviesbecca} and \citet{gadotti2019} observed the ionization cone in NGC 613 using [O \textsc{iii}]$\lambda$5007 images. In our analysis, we detected an extended soft X-ray emission around the nucleus of NGC 613. Such an extended emission around AGNs has been reported in the literature in many previous works (e.g. \citealt{young}, \citealt{hardcastle}, \citealt{bianchi}, \citealt{balmaverde}, \citealt{marinucci} and \citealt{greene}). However, different physical mechanisms have been proposed for such emission, depending on the observed properties. In some objects, for example, a spatial correspondence between the extended soft X-ray emission and the [O \textsc{iii}]$\lambda$5007 emission was observed (e.g. \citealt{young}, \citealt{bianchi}, \citealt{wang} and \citealt{balmaverde}). The most probable scenario for cases like these establishes that the extended X-ray emission comes from gas in the NLR photoionized by the central AGN. Other objects show a spatial correspondence between the extended X-ray emission and a radio jet or outflows from the AGN. In such cases, the most accepted scenarios involve inverse Compton scattering of radio photons by relativistic electrons in the jet (e.g. \citealt{hardcastle}) or shock heating due to the interaction between the jet (or outflow) and the interstellar medium (\citealt{ngc2110} and \citealt{greene}). In the case of NGC 613, Fig. \ref{chandramoles} shows a significant spatial correspondence between the ionization cone, represented by the [O \textsc{iii}]$\lambda$5007 emission, and the extended soft X-ray emission. Such a result suggests, together with the presence of some emission lines (Ne \textsc{ix}, Ne \textsc{x} Ly$_{\alpha}$ and Ne \textsc{x} Ly$_{cont}$), that there is a strong photoionization in the ionization cone. Considering the lower resolution of the X-ray data, the emission peak in soft X-rays is coincident with the position of [O \textsc{i}]$_{C}$ and of the N1 and N2 centres, but we can also see a strong emission towards the jet and the outflows detected in H$\alpha$ and [O \textsc{iii}]$\lambda$5007 (see Paper II for gas kinematics).

\subsection{Circumnuclear ring and nuclear spiral}\label{circumnuclear_ring}

As described in section \ref{introducao}, NGC 613 has a well-known circumnuclear ring of star formation. It is clearly seen in the image of Br$\gamma$ emission (presented in previous works and here in green in the composition of Fig. \ref{figextracaoespectro}), indicating, in this case, the presence of young stars that ionize the gas of the star-forming regions. \citet{boker} and \citet{falcon613} identified seven H \textsc{ii} regions along this circumnuclear ring. In this work, we could identify eight distinct H \textsc{ii} regions (one H \textsc{ii} region in addition to the other seven detected by \citealt{boker} and \citealt{falcon613}). The ring observed in the [Fe \textsc{ii}]$\lambda$16436 image (see Figs. 2 from \citealt{falcon613} and 13 from \citealt{audibert}) shows a higher granularity than in the Br$\gamma$ image. Considering such granularity and also the fact that the [Fe\textsc{ii}] emission is usually associated with shock heating from SNRs, we conclude that there may be SNRs distributed along the ring. \citet{boker} have already suggested the presence of SNRs along the cirumnuclear ring, based on the [Fe\textsc{ii}] emission from this area and \citet{falcon613} verified that the [Fe\textsc{ii}]/Br$\gamma$ ratio from the nuclear spectrum of NGC 613 also suggests shock heating from SNRs. We emphasize however that the higher granularity of the circumnuclear ring in the [Fe \textsc{ii}]$\lambda$16436 image is an additional argument supporting the presence of SNRs. The fact that the ring was observed both in hard X-ray (see Fig. \ref{raioxduros}) and radio \citep{hummel} reinforces the idea of SNRs along this region. Assuming that the ring is circular, we determined the parameters in Table \ref{anelparametros} and they are compatible with what was estimated by \citet{hummel}. Besides that, the radius is also compatible with the determination from \citet{audibert}.

\begin{table}
\centering
\caption{Parameters of the circumnuclear ring calculated from the images of Br$\gamma$ emission, [Fe \textsc{ii}]$\lambda$16436 and hard X-ray, assuming that the ring is circular.} \label{anelparametros}
\begin{tabular}{ccc}
\hline
Wavelength & Inclination (deg) & R (pc)       \\ \hline
Br$\gamma$ & 55 $\pm$ 5        & 225 $\pm$ 60 \\
{[}Fe \textsc{ii}{]}$\lambda$16436 & 61 $\pm$ 5        & 247 $\pm$ 60 \\
H$_2$$\lambda$21218  &  48 $\pm$ 5        & 223 $\pm$ 50\\
Hard X-Ray & 62 $\pm$ 9        & 301 $\pm$ 80 \\ 
Radio \citep{hummel} & 55  $\pm$ 5 &  $\sim$ 350 \\ 
Radio \citep{audibert} &      &   $\sim$ 300 \\ \hline
\end{tabular}
\end{table}

By calculating the emission-line ratios of the eight regions identified in the circumnuclear ring, we verified that all ratios are compatible with the ones of H \textsc{ii} regions, except regions  3 and 8, which have ratios compatible with the ones of LINERs (Fig. \ref{diagdiag}). That can be easily explained when we look at Fig. \ref{figextracaoespectro}, which shows that those regions are inside or close to the contour that delineates the ionization cone. The gas of those regions might be ionized inside the ionization cone, generating those emission-line ratios. Another possibility is that the emission-line ratios in these two regions result from a contamination  by the ionization cone emission. Region 8 that has the highest ionization degree might be also contaminated by the radio jet and outflows as we see in Fig. \ref{chandramoles}.

In order to study how the molecular emission is related to the whole scenario that we are analysing here, we used images from an ALMA data cube of CO(3-2) and the image of H$_2\lambda$21218 from SINFONI data cube (Fig. \ref{ANEL_NIR_ALMA}). The superposition criterion is described in Section \ref{matching}.  We see that neither [O \textsc{i}]$_{C}$ nor N2 are the centre of the nuclear spiral. And, by matching the circumnuclear ring, certainly, N2 is not the centre of the emission of H$_2\lambda$21218 and is not the centre of the nuclear spiral as it is represented in Fig. 1 of \citet{audibert}. What we see, as said previously, is V1 is the centre of the nuclear spiral.

As noted by \citet{audibert} the nuclear spiral flows from the ring towards the centre. The inner part of the spiral passes between N1 and N2, as we can see in Fig. \ref{HST_ALMA}(a) and in the proposed scheme in Fig. \ref{scenario}. When we compare this structure with the compositions of the \textit{HST} filters in Figs. \ref{hst_tot} and \ref{GMOS_HST}(d), we can see that there is a significant reddening and obscuration by dust between N1 and N2. This may indicate that the spiral is probably bringing dust and gas to the centre of the galaxy. Since the spiral is located between N1 and N2 and we cannot see a clear double structure in SINFONI data, together with the fact that we did not detect any difference in the stellar populations of N1 and N2 (Paper II), we may have only one source in the centre of NGC 613. In that case, the dust extinction caused by the nuclear spiral results in an apparent division of the central source in the components N1 and N2.

Fig. \ref{HST_ALMA}(b) shows, in red, areas affected by dust extinction. We clearly see a connection between these areas and the molecular ring. This reinforces the hypothesis of the feeding of the circumnuclear ring by the bar proposed by \citet{boker} and \citet{audibert}. The morphology also suggests little connection between the feeding coming from the bar and the nuclear spiral. We observe the same feature in H$\alpha$ image in Fig. 10 from \citet{gadotti2019}.

\subsection{The global scenario for the central region}\label{cenariogeraldiscuss}

We drew a scheme to define the scenario in the central 2 arcsec, taking into account all the phenomena that we described in this paper (see Fig. \ref{scenario}). This scheme is what we defined here as being the scenario that explains the coexistence of the multiple structures studied in this work. The nuclear spiral (in blue) passes between N1 and N2 and may bring gas and dust from the circumnuclear region. Regions N1 and N2 are represented by the two grey circles, corresponding to the two components of the double stellar nucleus as appears in the \textit{HST} images. Since they are separated by a stream of dust and the nuclear spiral, we are not entirely sure if they are separated regions. There are, therefore, two hypotheses to explain them: they are probably part of one central extended structure (as seen in Fig. \ref{SINFONIN1N2}) and the separation might be an obscuration effect caused by the dust. The second hypothesis assumes that they are two separated structures orbiting the central AGN in the plane of the observation, since we did not detect any difference of velocities in both regions (see Paper II). In Paper II we will resume this discussion.

V1 and V2, in Fig. \ref{scenario}, are the two variable sources. We believe that V1 is the AGN, since it is in the centre of the nuclear spiral and it is a very strong point-like source in the \textit{HST} images. If this is correct, the AGN has variable activity that was also proposed by \citet{audibert} when looking for possible fossil molecular outflows. N1 and N2 might be ionized by the central AGN. [O \textsc{i}]$_{C}$ is not compatible with the position of V1 and this is probably due to differential extinction together with emission and reflection of the [O \textsc{i}]$\lambda$6300 emission in the ionization cone. The position of V2 suggests that it might be a supernova, since it is inside a cluster of young stellar populations (that is N1, see Paper II).

\section{Conclusions}\label{secconclusion}

Multiwavelength analysis has shown that the galaxy NGC 613 has a rich nuclear environment. The study of data cubes from telescopes in the NIR, optical, X-ray, and radio bands, besides \textit{HST} images, led us to the following findings:

$\bullet$ In the optical band, the central region of NGC 613 is characterized by an apparent double stellar nucleus. We called the two stellar nuclei as N1 and N2. The least brightest nucleus, as seen in the \textit{HST} images, coincides, within the errors, with the emission of [O \textsc{i}]$\lambda$6300,  whose centre was defined as [O \textsc{i}]$_{C}$ in this work. The brightest nucleus as seen in the \textit{HST} image was defined as N2. The separation between the two stellar components, as seen in the \textit{HST} images, is $\sim$ 94 pc ($\sim$ 0.74 arcsec). 

$\bullet$ The spectrum extracted from N1 has emission-lines ratios compatible with the ones of LINERs, presenting fairly broad forbidden lines (FWHM $\sim$ 665 km s$^{-1}$). On the other hand, the spectrum from N2 has also emission-line ratios of LINERs, but with slightly higher ionization degree than N1.

$\bullet$ By comparing the images in the filter \textit{F 814 W} from the \textit{HST} observed in 2001 and in 2018, we detected evidence of variability in a point between N1 and N2, whose distance from [O \textsc{i}]$_{C}$ is 0.24 arcsec. In 2001 this source was not detected and in 2018 it was clearly visible. The position of [O \textsc{i}]$_{C}$ and of this variable source are not compatible, but they are very close to each other. Since this variable source is point-like in the \textit{HST} images and is the centre of all nuclear structure, it might be the central variable AGN. Therefore, the shift between the centre of the [O \textsc{i}]$\lambda$6300 emission (typically associated with partial ionization regions in AGNs), [O \textsc{i}]$_{C}$, and this variable central source (the AGN in this case) might be due to scattering of the AGN emission and emission of the ionization cone, since the shift is towards the ionization cone.

$\bullet$ When we compare the optical data from \textit{HST} and GMOS, observed in 2001 and 2015, respectively, we verify that N1 also suffered a variability in brightness. In 2001 this source was fainter than in 2015. Considering the time interval and also that N1 has young stellar populations, this might be an evidence of a supernova.

$\bullet$ We found extended soft X-ray emission, closely associated with the ionization cone, seen in [O \textsc{iii}]$\lambda$5007. There is also an excess of this emission towards the radio jet and gas outflows, which might be disturbing the circumnuclear ring.

$\bullet$ The hard X-ray emission, besides having a strong central component, also presents a circumnuclear structure in the form of a ring that has geometric parameters similar to those seen in Br$\gamma$, [Fe \textsc{ii}]$\lambda$16436, and H$_2\lambda$21218 and radio images. This is likely originated from SNRs associated with star formation in the ring. The high granularity of the [Fe \textsc{ii}]$\lambda$16436 image, when compared with the Br$\gamma$ image, reinforces this hypothesis.

$\bullet$ The profile of the central hard X-ray emission is slightly broader than the PSF by 0.33 arcsec. We interpret this as possibly due to circumnuclear scattering.

$\bullet$ From the optical (SIFS) and NIR (SINFONI) data we identified 10 H \textsc{ii} regions. Eight of them are part of the circumnuclear ring (already observed in other works), while the other two are further away.

$\bullet$ There are at least three H \textsc{ii} regions in the circumnuclear ring whose emission-line ratios are affected by emission from the ionization cone or partially ionized by the central source. As a consequence, they present spectra that resemble the ones of LINERs.

$\bullet$ We confirm, by analysing the \textit{HST} data together with the CO(3-2) image from ALMA, that the molecular gas ring is being fed by the bar along two arms. The nuclear spiral, which passes between N1 and N2, might be bringing dust and gas to the centre, causing the obscuration. Also it might be dividing an extended central stellar structure in two, that is the double stellar nucleus that we see in the optical band, due to dust obscuration.

\section*{Acknowledgements}

This work is based on observations obtained at the Gemini Observatory (processed using the Gemini \textsc{iraf} package), which is operated by the Association of Universities for Research in Astronomy, Inc., under a cooperative agreement with the NSF on behalf of the Gemini partnership: the National Science Foundation (United States), the National Research Council (Canada), CONICYT (Chile), the Australian Research Council (Australia), Minist\'erio da Ci\^encia, Tecnologia e Inova\c{c}\~ao (Brazil), and Ministerio de Ciencia, Tecnolog\'ia e Innovaci\'on Productiva (Argentina). This work is also based on observations made with the NASA/ESA \textit{Hubble Space Telescope}, obtained from the Data Archive at the Space Telescope Science Institute, which is operated by the Association of Universities for Research in Astronomy, Inc., under NASA contract NAS 5-26555. This research has also made use of the NASA/IPAC Extragalactic Database (NED), which is operated by the Jet Propulsion Laboratory, California Institute of Technology, under contract with the National Aeronautics and Space Administration. This research has also made use of data obtained from the \textit{Chandra} Data Archive and the \textit{Chandra} Source Catalog, and software provided by the \textit{Chandra} X-ray Center (CXC) in the application packages \textsc{ciao}, \textsc{chips}, and \textsc{sherpa} and used data from observations collected at the European Southern Observatory under ESO programme 076.B-0646(A). This paper also makes use of the following ALMA data: ADS/JAO.ALMA 2015.1.00404.S. ALMA is a partnership of ESO (representing its member states), NSF (USA) and NINS (Japan), together with NRC (Canada), MOST and ASIAA (Taiwan), and KASI (Republic of Korea), in cooperation with the Republic of Chile. The Joint ALMA Observatory is operated by ESO, AUI/NRAO and NAOJ. We thank CNPq (Conselho Nacional de Desenvolvimento Cient\'ifico e Tecnol\'ogico), under grant 141766/2016-6, and FAPESP (Funda\c{c}\~ao de Amparo \`a Pesquisa do Estado de S\~ao Paulo), under grant 2011/51680-6, for supporting this work. We also thank professor Giuseppina Fabbiano for offering suggestions on \textit{Chandra} data cube analysis and Dr. Roderik Overzier for reading and revising this article.




\bibliographystyle{mnras}
\bibliography{references} 



\appendix

\section{Spectra of the observed regions and emission-line decompositions}\label{espectros_analise}

\begin{figure*}
\begin{center}
   \includegraphics[scale=0.31]{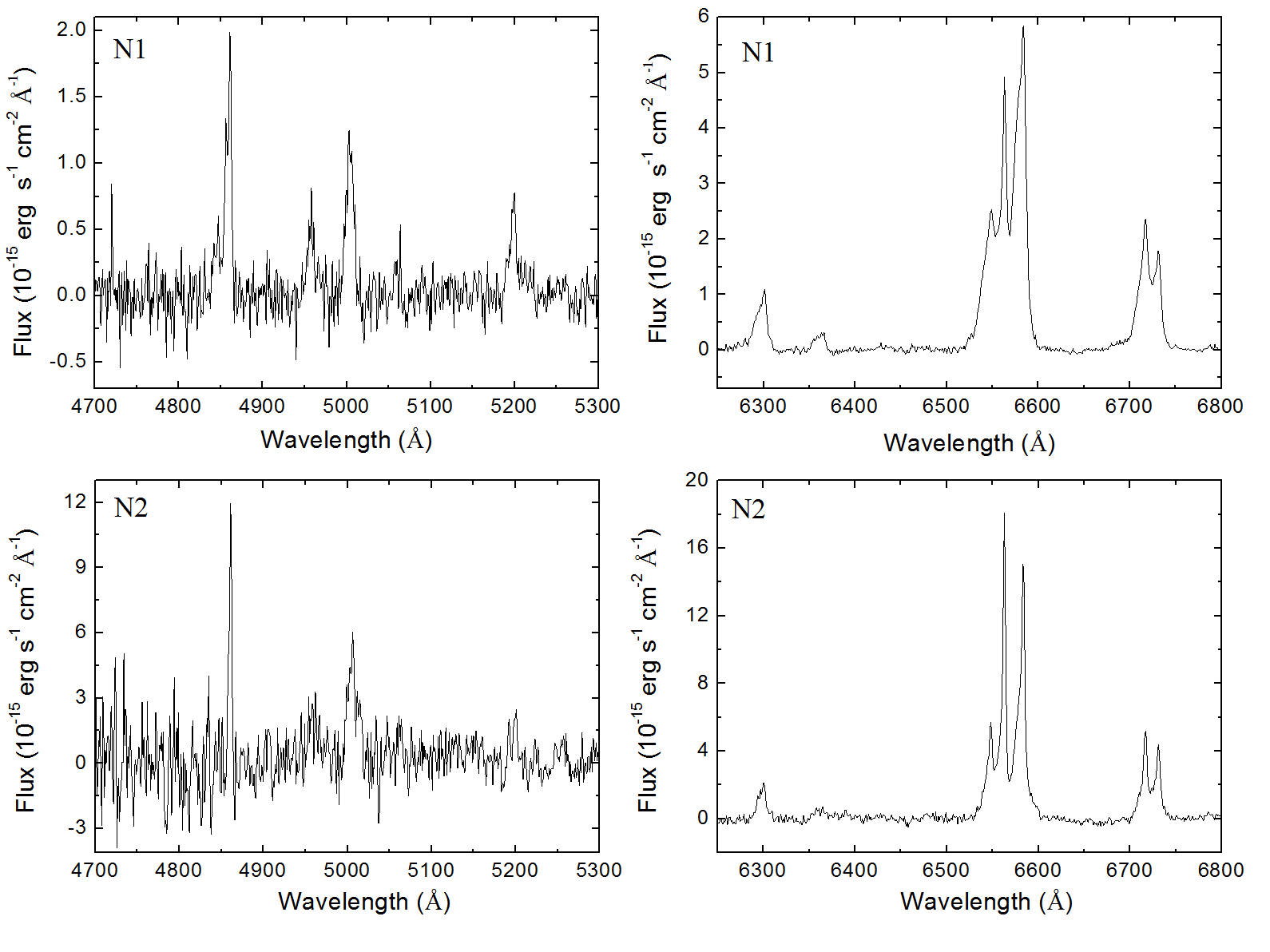}
  \caption{Blue and red intervals of the spectra of regions N1 and N2 extracted from the GMOS data cube. \label{espectrosN1N2}}
\end{center}
\end{figure*}

As explained in section \ref{sec_razaodelinhas}, we extracted a spectrum of each identified region in Fig. \ref{figextracaoespectro}. The extraction was performed using circular areas, whose diameter was taken as the FWHM of each instrument: for regions 1 to 10, we used the SIFS data cube and, for N1 and N2 regions, we used the GMOS data cube. Figs. \ref{espectrosN1N2}, \ref{espectros12345} and \ref{espectros678910} show the blue and red parts of the extracted spectra of each region.

One can notice that the spectra from regions N1, N2, 2, 3, 7, and 8 have blended emission lines. In order to calculate the emission-line ratios of each region, it was necessary to decompose the blended lines in two Gaussian sets. First, we determined an uncertainty to the flux associated with each wavelength of the spectrum. For this, we calculated the standard deviation  of the values in a specific wavelength range, without any emission line, of the extracted spectra. These values of standard deviation were taken as the uncertainty of the fluxes in the average wavelengths of the ranges used in this estimate. Then, we made an interpolation of the values, in order to obtain a value of uncertainty to each wavelength of the spectrum. 

\begin{figure*}
\begin{center}
   \includegraphics[scale=0.3]{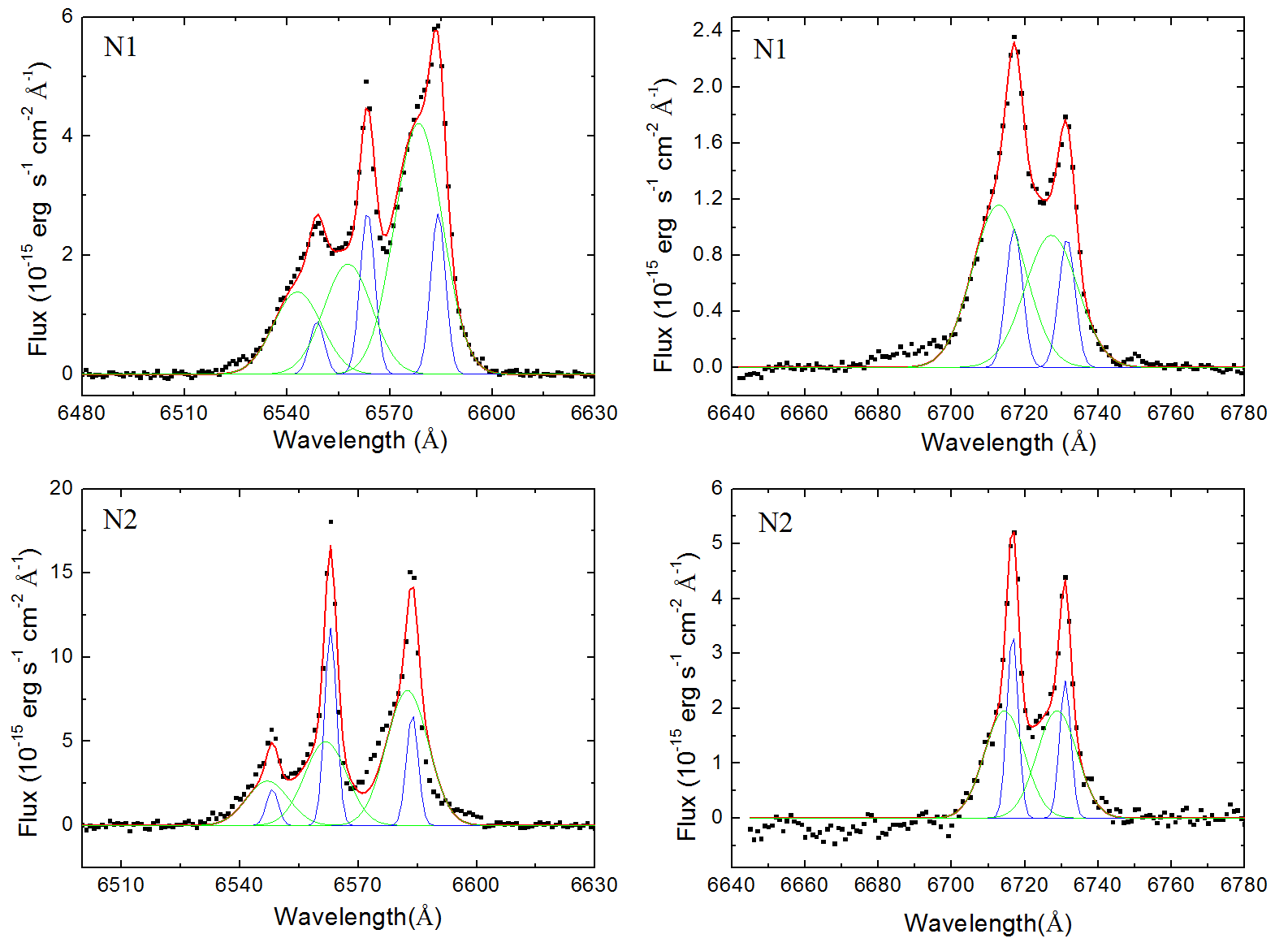}
  \caption{Decomposition of the blended H$\alpha$ + [N \textsc{ii}]$\lambda\lambda$6548, 6584 and [S \textsc{ii}]$\lambda\lambda$6716, 6731 lines in order to calculate the emission-line ratios of regions N1 and N2 (indicated in each panel) of the GMOS data cube. The blue and green Gaussians are the narrow components of the fitted lines, the red curve represents the total fit and the points are the observed data. \label{decomposicaoN1N2}}
\end{center}
\end{figure*}

\begin{figure*}
\begin{center}
   \includegraphics[scale=0.30]{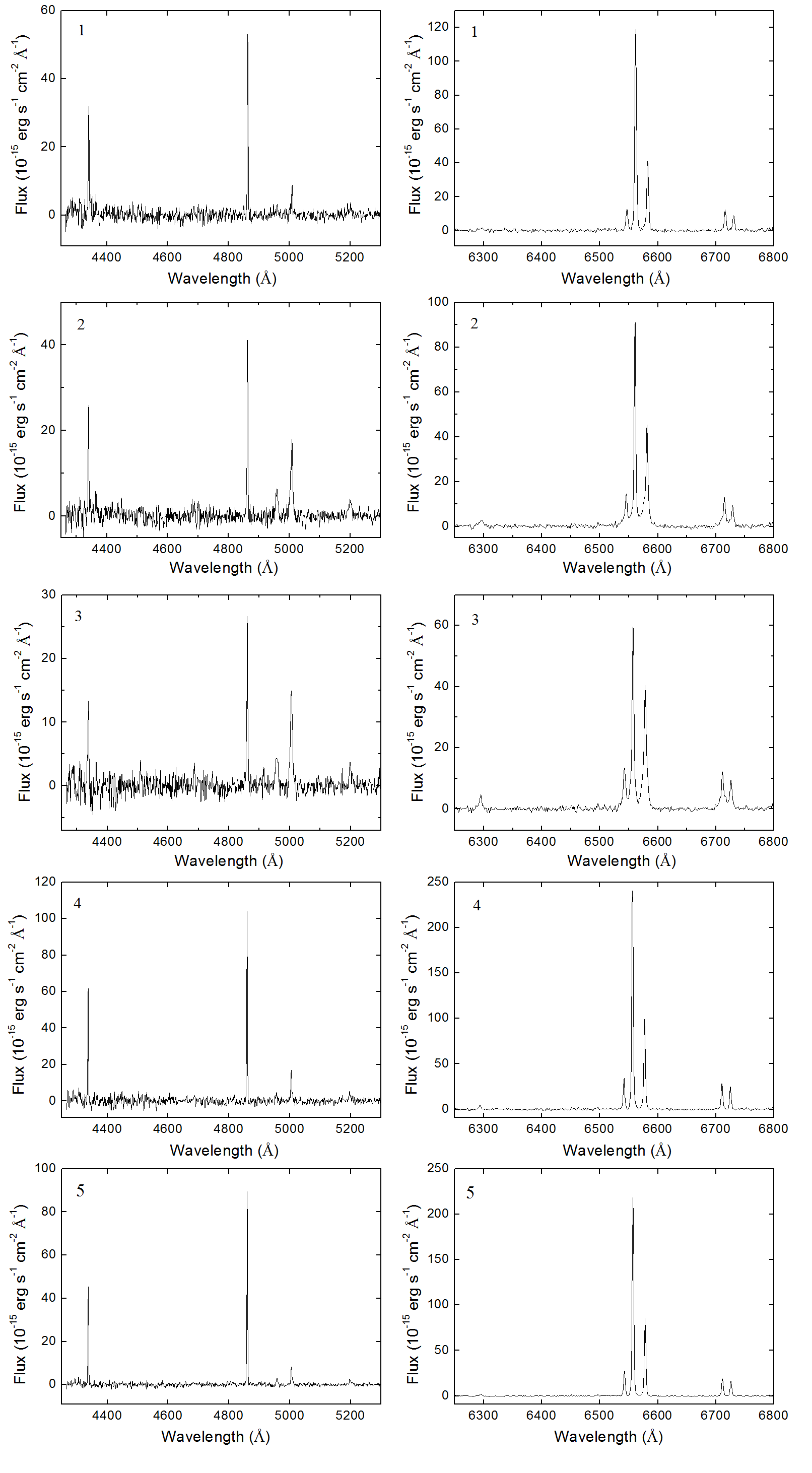}
  \caption{Blue and red intervals of the spectra of regions 1, 2, 3, 4, and 5 extracted from the SIFS data cube. \label{espectros12345}}
\end{center}
\end{figure*}

\begin{figure*}
\begin{center}
   \includegraphics[scale=0.30]{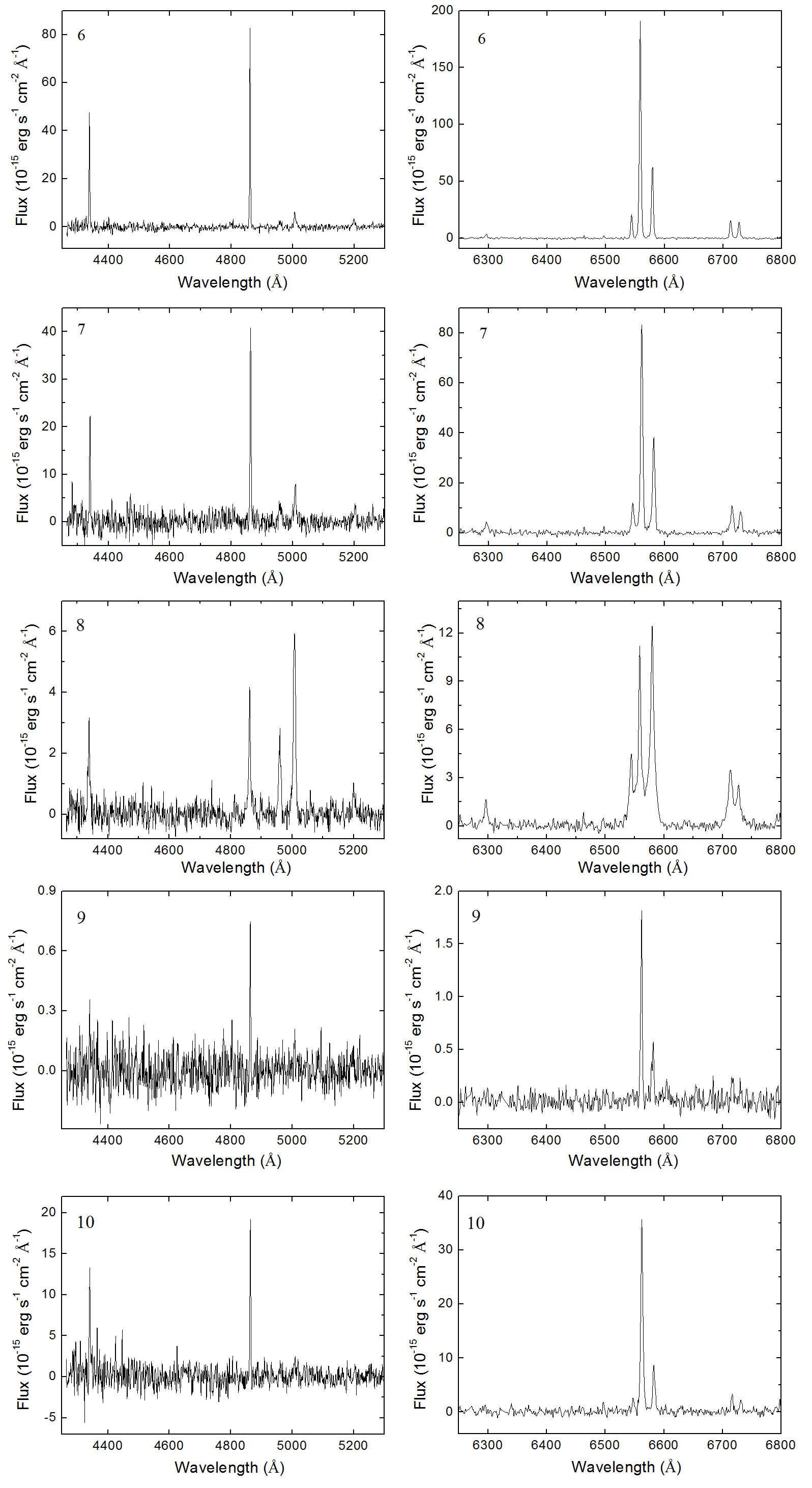}
  \caption{Blue and red intervals of the spectra of regions 6, 7, 8, 9, and 10 extracted from the SIFS data cube. \label{espectros678910}}
\end{center}
\end{figure*}

\begin{figure*}
\begin{center}
   \includegraphics[scale=0.30]{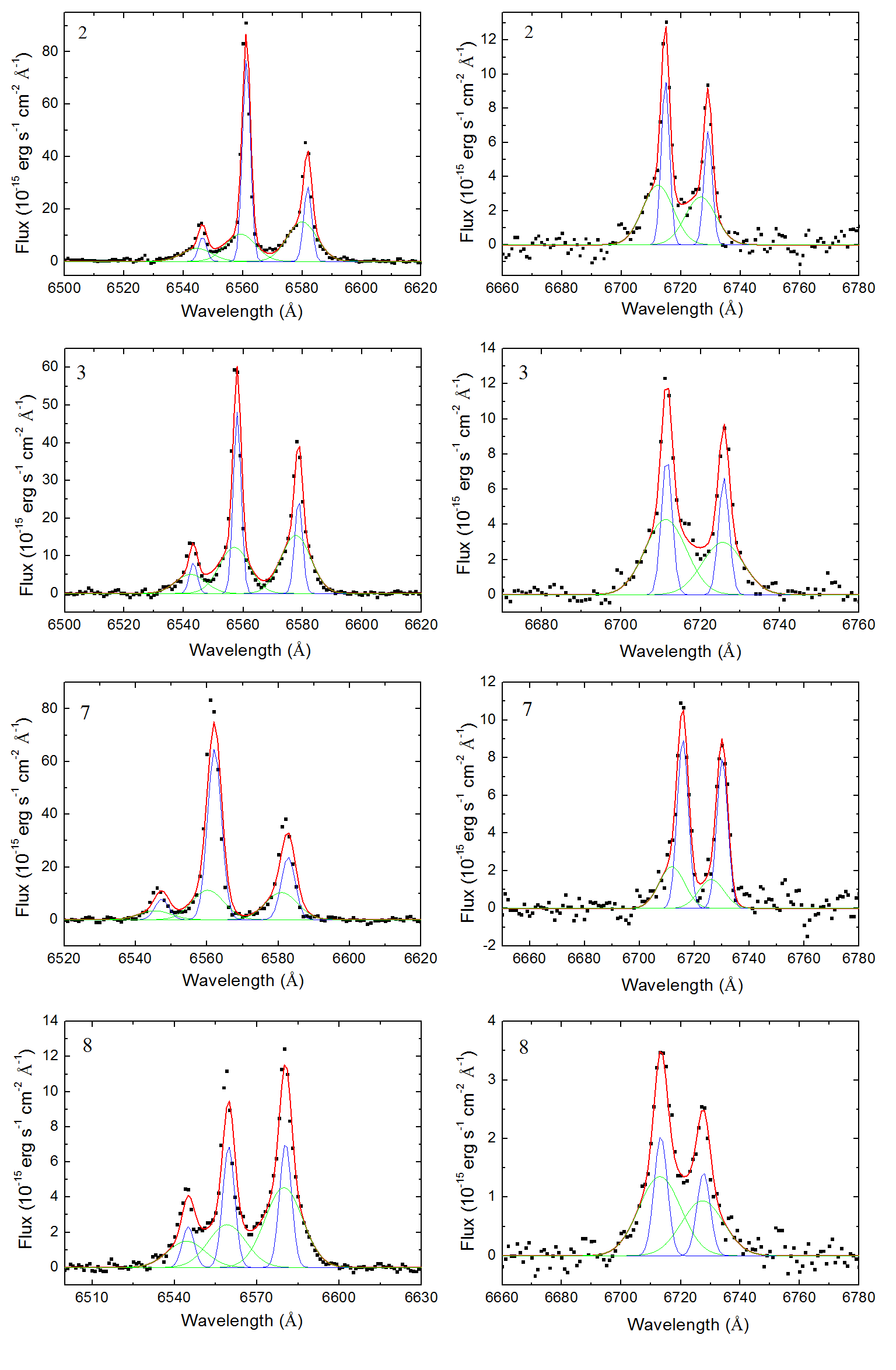}
  \caption{Decomposition of the blended H$\alpha$ + [N \textsc{ii}]$\lambda\lambda$6548, 6584 and [S \textsc{ii}]$\lambda\lambda$6716, 6731 lines, in order to calculate the emission-line ratios of regions 2, 3, 7, and 8 (indicated in each panel) of the GMOS data cube. The blue and green Gaussian fits are the narrow components of the fitted lines, the red curve represents the total fit and the points are the observed data. \label{decomposicao1378}}
\end{center}
\end{figure*}

The first Gaussian fit involved the lines [S \textsc{ii}]$\lambda \lambda$6716;6731. Each line was fitted by a sum of two Gaussian functions: the green Gaussians set and the blue Gaussians set in Figs. \ref{decomposicaoN1N2} and \ref{decomposicao1378}. For each set, we adopted a specific width and velocity (free parameters of the fits). In other words, each line of [S \textsc{ii}] was fitted by a sum of two Gaussians functions (green+blue), each one with a specific width and velocity. 

It is important to mention that the maximum and minimum values admitted for the ratio of the integrated fluxes of the [S \textsc{ii}]$\lambda$6716 and [S \textsc{ii}]$\lambda$6731 Gaussians, in each one of the two sets, was 1.44 and 0.44, respectively, as theoretically established by \citet{osterbrock}.

After that, we fitted the lines [N \textsc{ii}]$\lambda\lambda$6548;6584 + H$\alpha$ as a sum of two Gaussians for each line, which resulted in a set of three green Gaussians and a set of three blue Gaussians in Figs. \ref{decomposicaoN1N2} and \ref{decomposicao1378}. We assumed that, in each set, the Gaussians have the same width and redshifts as the corresponding Gaussians of the [S \textsc{ii}]$\lambda\lambda$6716;6731 emission lines. In other words, the [S \textsc{ii}]$\lambda\lambda$6716;6731 lines were taken as an empiric template to these fits. 

We determined, then, for each spectrum, the value of the H$\alpha$/H$\beta$ ratio (Balmer decrement). By using this ratio and also the extinction law determined by \citet{cardelli}, we applied the interstellar extinction correction to each spectra. Then, we calculated the integrated fluxes of the H$\beta$, [O \textsc{iii}]$\lambda$5007, [O \textsc{i}]$\lambda$6300, [N \textsc{ii}]$\lambda \lambda$6548;6584, H$\alpha$ e [S \textsc{ii}]$\lambda\lambda$6716;6731 lines of the corrected spectra, after applying the previous process of decomposition when it was necessary. The integrated fluxes of the non-blended lines were calculated by a direct integration. From these values we calculated the emission-line ratios as shown in section \ref{sec_razaodelinhas}.

The uncertainty of the integrated flux of each blended line was obtained from the propagation of the uncertainties determined for the parameters of the Gaussian fits. On the other hand, the uncertainty of the integrated flux of each non-blended line was taken considering the different ranges of integration. This procedure was adopted since such uncertainties take into account not only the spectral noise but also possible irregularities in the spectrum caused by inaccuracies in the process of stellar continuum subtraction.

\bsp	
\label{lastpage}
\end{document}